\documentclass{emulateapj}

\usepackage{natbib}
\usepackage{hyperref}
\usepackage{breakurl}

\newcommand{\msol}{{\rm M}_{\sun}}
\newcommand{\rsol}{{\rm R}_{\sun}}

\newcommand{\kms}{\rm km\ s^{-1}}

\begin{document}

\shorttitle{Four eclipsing mid M-dwarf systems}
\shortauthors{Irwin et al.}

\title{Four new eclipsing mid M-dwarf systems from the New Luyten
Two Tenths catalog}

\author{Jonathan~M.~Irwin, David~Charbonneau, Gilbert~A.~Esquerdo,
  David~W.~Latham, Jennifer~G.~Winters}
\affil{Harvard-Smithsonian Center for Astrophysics, 60 Garden St.,
  Cambridge, MA 02138, USA}

\author{Jason~A.~Dittmann, Elisabeth~R.~Newton}
\affil{Kavli Institute for Astrophysics and Space Research, Massachusetts Institute of Technology, 77 Massachusetts Avenue, Cambridge, MA 02139, USA}

\author{Zachory~K.~Berta-Thompson}
\affil{Department of Astrophysical and Planetary Sciences, University of Colorado, Boulder, CO 80309, USA}

\author{Perry~Berlind and Michael~L.~Calkins}
\affil{Fred Lawrence Whipple Observatory, Smithsonian Astrophysical Observatory, 670 Mount Hopkins Road,
  Amado, AZ 85645, USA}

\begin{abstract}
Using data from the MEarth-North and MEarth-South transit surveys, we
present the detection of eclipses in four mid M-dwarf systems:
LP~107-25, LP~261-75, LP~796-24, and LP~991-15.  Combining the MEarth
photometry with spectroscopic follow-up observations, we show that
LP~107-25 and LP~796-24 are short-period (1.388 and 0.523 day,
respectively) eclipsing binaries in triple-lined systems with
substantial third light contamination from distant companions.
LP~261-75 is a short-period (1.882 day) single-lined system consisting
of a mid M-dwarf eclipsed by a probable brown dwarf secondary, with
another distant visual brown dwarf companion.  LP~991-15 is a
long-period (29.3 day) double-lined eclipsing binary on an eccentric
orbit with a geometry which produces only primary eclipses.  A
spectroscopic orbit is given for LP~991-15, and initial orbits
for LP~107-25 and LP~261-75.
\end{abstract}


\section{Introduction}

Eclipsing binaries are important astrophysical tools because they are
able to provide largely model-independent, precise measurements of
stellar masses and radii.  Observations of the best-characterized
examples reveal a systematic tendency of theoretical stellar evolution
models to underpredict the radii of main sequence stars with
convective outer envelopes (e.g.,
\citealt{1991A&ARv...3...91A,2010A&ARv..18...67T}).

Due to the special geometric configuration required for a
spectroscopic binary to eclipse, such systems are rare.
Observations are particularly sparse for fully convective M-dwarfs 
(stars with masses below approximately $0.35\ \msol$; e.g.,
\citealt{1997A&A...327.1039C}), and while recent observational
progress has begun to fill in the parameter space between $0.2\ \msol$
and $0.35\ \msol$, there are still very few systems containing
components below $0.2\ \msol$ with precisely measured parameters
(e.g., \citealt{2013MNRAS.431.3240N,2017ApJ...836..124D}).

We operate the MEarth project, an all-sky survey using two robotic
telescope arrays to search for transiting planets orbiting fully
convective M-dwarfs within $33\ {\rm pc}$ by obtaining high-cadence 
time series differential photometry \citep{2008PASP..120..317N}.  This
survey is also highly sensitive to eclipsing binaries, which present
much larger photometric signal sizes than transiting planets, and has
been optimized for efficient recovery of objects with long orbital
periods.

In previous papers, we have presented three eclipsing stellar systems
\citep{2009ApJ...701.1436I,2011ApJ...742..123I,2017ApJ...836..124D}
and one brown dwarf system \citep{2010ApJ...718.1353I} detected with
MEarth.  This paper presents details of four additional eclipsing
systems detected over the same time period.  Three are stellar systems
(LP~107-25, LP~796-24 and LP~991-15), for which we report initial
observations and characterization, but several concerns must be
addressed before masses and radii of the components can be
determined at the level required to test stellar models.  The fourth
(LP~261-75) is a single-lined system with a probable brown dwarf
companion.

\section{Target selection and properties}

Target selection for MEarth-North is described in detail in
\citet{2008PASP..120..317N}, and for MEarth-South in
\citet{2015csss...18..767I}.  All four targets presented here were
selected for observation based on photometric distance estimates
placing them within 33 pc, a volume limit inherited from the work
  of \citet{2005AJ....130.1680L}, upon which our original target
  selection was based.  These distances were underestimated in three
cases due to these targets being unresolved multiples in the original
photometry.  Overluminosity results in a larger effective volume limit
for such systems in MEarth when using photometric distance estimates,
enhancing the detection rate.

This factor combined with the greatly improved availability of
astrometric parallaxes for the MEarth sample in recent years, and the
continuing demand to push to smaller planet sizes, has resulted in
reprioritization of the MEarth target list and elimination or
deprioritization of the majority of these more distant sources.  These
changes were implemented during the 2016--2017 period.  While this is
very beneficial for planet detection it has likely considerably
impacted detection of new eclipsing binaries, so in this publication
we provide details of the final accumulated set of systems detected
prior to the completion of these changes which were not published
individually.

In Table \ref{photparams}, we summarize the photometric and
astrometric properties of the four systems gathered from the
literature: right ascension and declination $\alpha$, $\delta$; proper
motions $\mu_\alpha$, $\mu_\delta$; astrometric parallax $\pi_{\rm
  trig}$; GAIA $G$, $G_{\rm BP}$ and $G_{\rm RP}$ photometry; infrared
magnitudes $J_{\rm 2MASS}$, $H_{\rm 2MASS}$, $K_{\rm 2MASS}$ from
2MASS \citep{2006AJ....131.1163S}; and spectral type (where
available).  We use the identifiers given in the NLTT throughout this
work, but for LP~107-25 this identifier does not appear to have been
used previously in the literature, so we also provide 2MASS
identifiers in the table as an alternative.  GAIA parameters are from
Data Release 2 (DR2), which was released during the final stages of
preparation of the manuscript.  We have updated the parallaxes, but
the positions and proper motions given in the table are the original
ones assumed during the analysis.

\begin{deluxetable*}{lrrrr}
\tablecaption{\label{photparams} Summary of the photometric and
  astrometric properties of the four systems.}
\tablecolumns{5}

\tablehead{
\colhead{Parameter}         & \colhead{LP~107-25}  & \colhead{LP~261-75}  & \colhead{LP~796-24}  & \colhead{LP~991-15}
}

\startdata
2MASS identifier & J21280940+6321013 & J09510459+3558098 & J13004029-2010434 & J01234181-3833496 \\
\\
$\alpha_{\rm ICRS, 2000.0}$ & 21:28:09.40             & 09:51:04.58           & 13:00:40.26           & 01:23:41.84 \\
$\delta_{\rm ICRS, 2000.0}$ & $+$63:21:01.4           & $+$35:58:09.5         & $-$20:10:43.8         & $-$38:33:49.6 \\
$\mu_\alpha$ (arcsec/yr)    & $ 0.036 \pm 0.008$      & $-0.106 \pm 0.008$    & $-0.292 \pm 0.002$    & $0.243$ \\
$\mu_\delta$ (arcsec/yr)    & $ 0.189 \pm 0.008$      & $-0.171 \pm 0.008$    & $-0.156 \pm 0.002$    & $0.000$ \\
Source                      & 1                       & 1                     & 2                     & 2MASS/NLTT \\
\\
$\pi_{\rm trig}$ (arcsec)   & $0.021033 \pm 0.000094$ & $0.02945 \pm 0.00014$ & $0.02529 \pm 0.00012$ & $0.031633 \pm 0.000064$ \\
$G$                         & $12.749$                & $13.833$              & $13.980$              & $13.736$ \\
$G_{\rm BP}$                & $14.163$                & $15.635$              & $15.840$              & $15.541$ \\
$G_{\rm RP}$                & $11.586$                & $12.541$              & $12.692$              & $12.455$ \\
Source                      & 3                       & 3                     & 3                     & 3 \\
\\
$J_{\rm 2MASS}$             & $ 9.988 \pm 0.024$      & $10.577 \pm 0.021$    & $10.814 \pm 0.022$    & $10.620 \pm 0.026$ \\
$H_{\rm 2MASS}$             & $ 9.406 \pm 0.028$      & $ 9.960 \pm 0.019$    & $10.205 \pm 0.025$    & $10.059 \pm 0.024$ \\
$K_{\rm 2MASS}$             & $ 9.164 \pm 0.022$      & $ 9.690 \pm 0.019$    & $ 9.918 \pm 0.021$    & $ 9.749 \pm 0.023$ \\
\\
Spectral type               & \ldots                  & M4.5                  & M4.5                  & M4.5 \\
Source                      & \ldots                  & 4                     & 5                     & 5 \\
\enddata

\tablerefs{(1) \citet{2005AJ....129.1483L}, (2) \citet{2017A+A...600L...4A}, (3) \citet{2018arXiv180409365G}, (4) \citet{2006PASP..118..671R}, (5) \citet{2003AJ....126.3007R}.}

\end{deluxetable*}

\section{MEarth photometric observations}

The MEarth data themselves, data reduction, and analysis
methods have been described in detail in previous papers
\citep{2011ApJ...727...56I,2011ApJ...742..123I,2012AJ....144..145B,2016ApJ...821...93N}.
The objects presented here were detected during the 2011--2017
observing seasons, during which time the configuration of both
instruments was relatively stable, with all observations taken using
the same RG715 filter bandpass.

Eclipses in LP~107-25 were detected during a search for photometric
rotation periods performed at the end of the 2011--2013 observing
period on approximately weekly cadence data intended for astrometry.
The other three systems were all detected by MEarth's real-time
trigger based on single events in 2014 June (LP~991-15), 2014 December
(LP~796-24), and 2017 June (LP~261-75).

In all four cases, after the initial detection was made additional
photometric observations were gathered for follow-up, both during the
eclipse windows at high cadence using back-to-back exposures, and at
our standard 30 minute cadence between eclipse windows to search for
any out-of-eclipse variations.  In these mid M-dwarf systems the
out-of-eclipse variations are usually dominated by rotational
modulation (presumed to be due to starspots) rather than effects
intrinsic to close binary stars such as ellipsoidal variation or
reflection, but are important for modeling.

MEarth light curves require some pre-processing to remove bad data
prior to use.  For this publication we simply filter out data points
taken through heavy clouds by requiring the magnitude zero point for
the image was no more than $0.5\ {\rm mag}$ higher than a running
average, where this average was computed by outlier resistant fitting
of straight line segments to the magnitude zero point as a function of
time for images taken with a stable instrument configuration (the same
``instrument version number'' in the light curve table).  We note that
the throughput has evolved substantially over the time period of
observations presented here and shows a large jump at the 2016 summer
shutdown when the telescope optics were cleaned, and we consider
points taken before and after this time to be different instrument
configurations for the purpose of this analysis.  A few data points
with large pointing errors due to target acquisition problems were
also discarded.  The light curve data are given in an electronic
table.  This table provides the original light curve data used as
input to the models, so the corrections described in \S
\ref{lc_nuisance_sect} have not been applied.

\begin{figure*}
\centering
\includegraphics[width=1.74in]{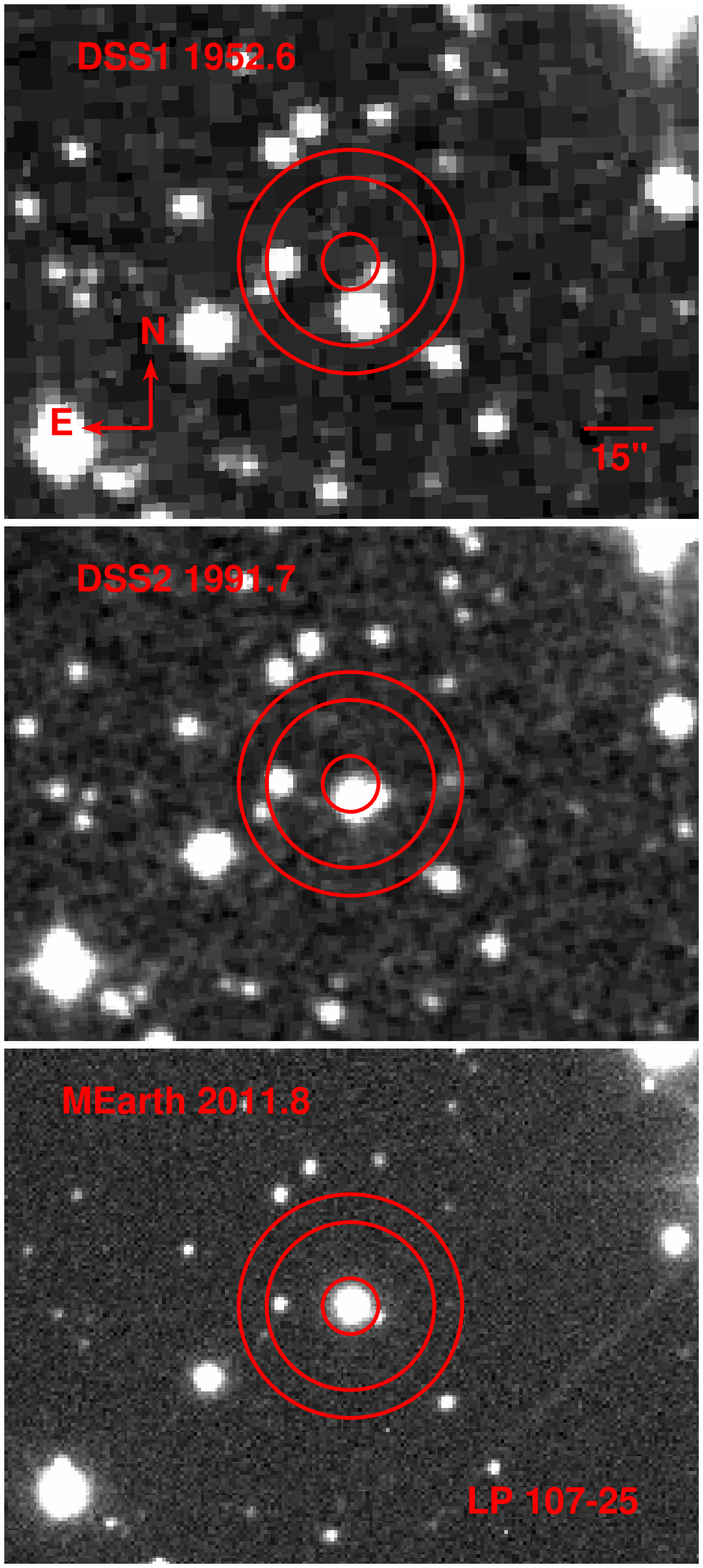}
\includegraphics[width=1.74in]{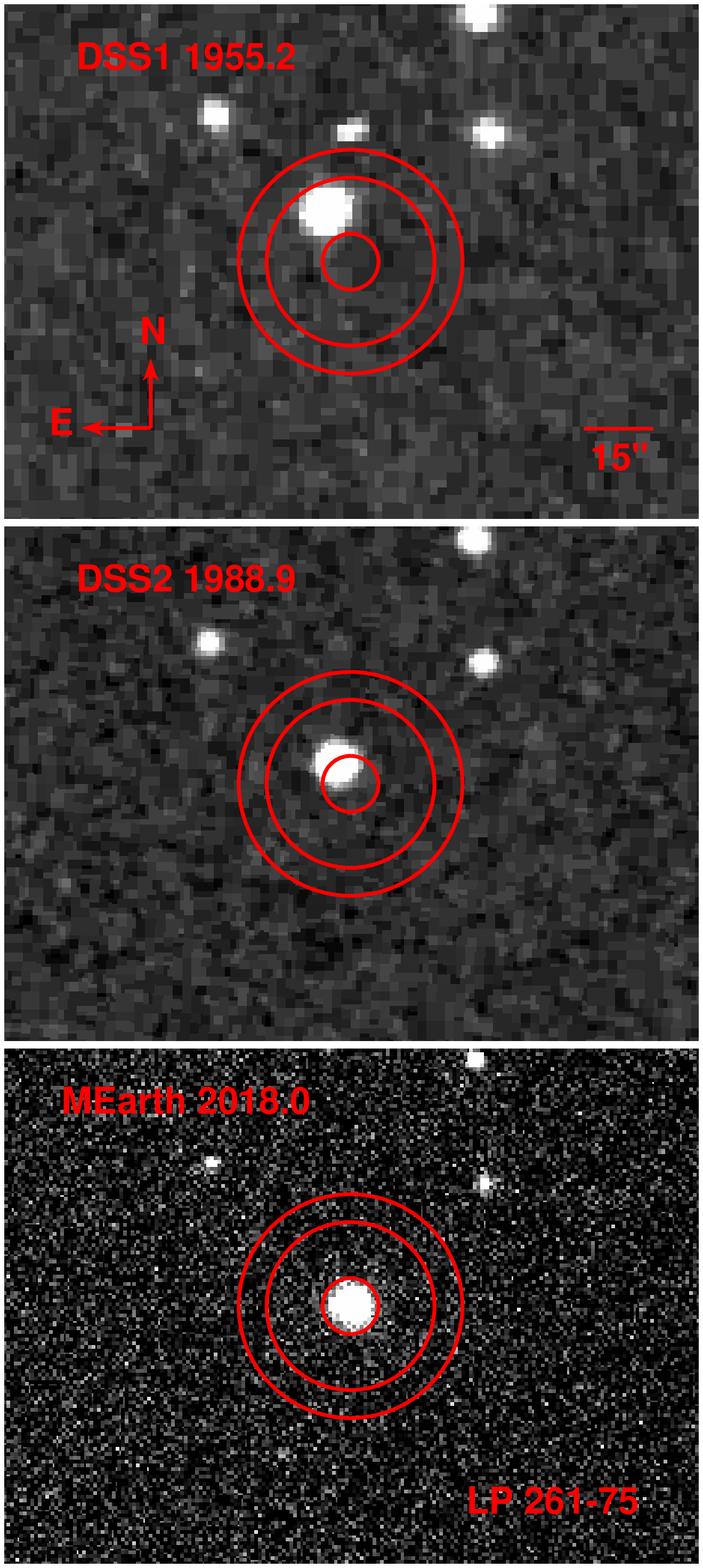}
\includegraphics[width=1.74in]{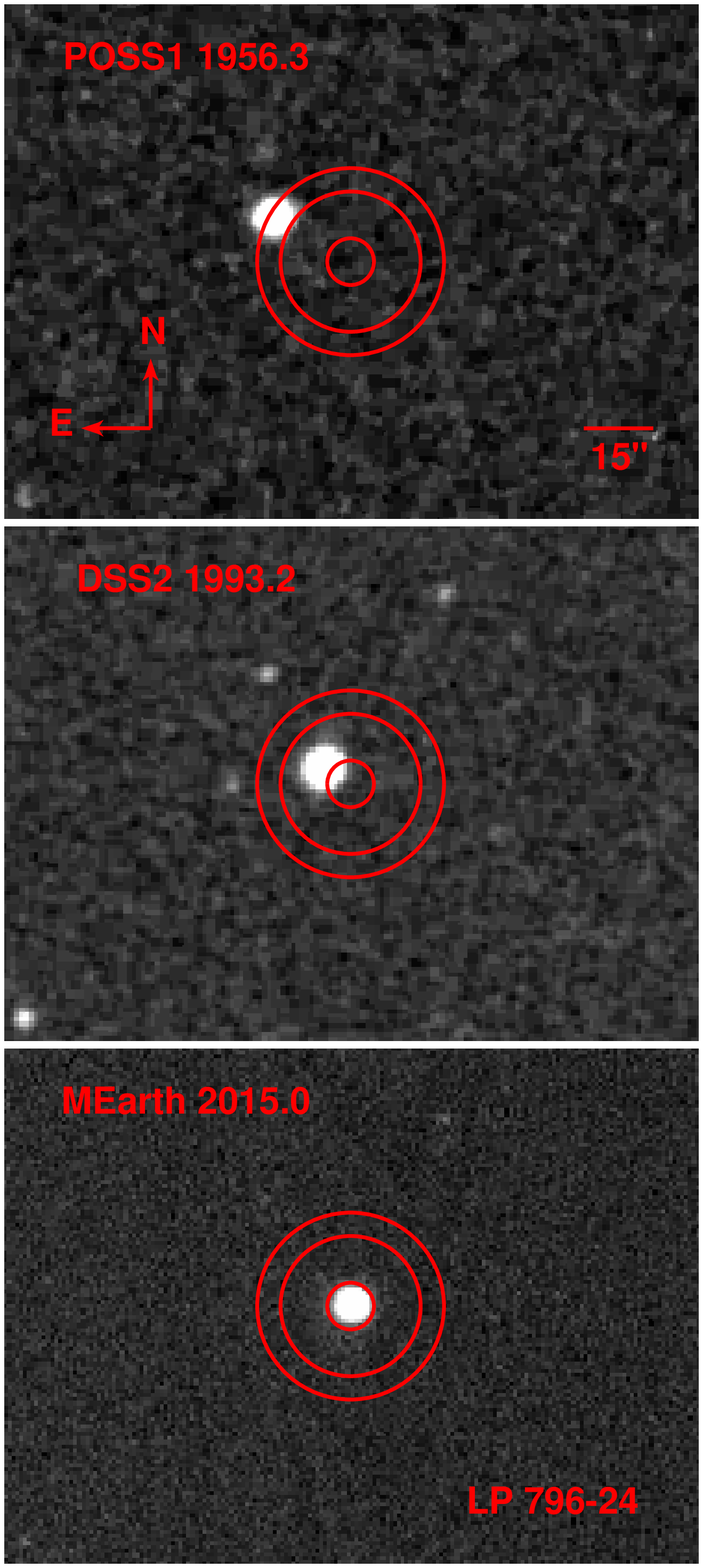}
\includegraphics[width=1.74in]{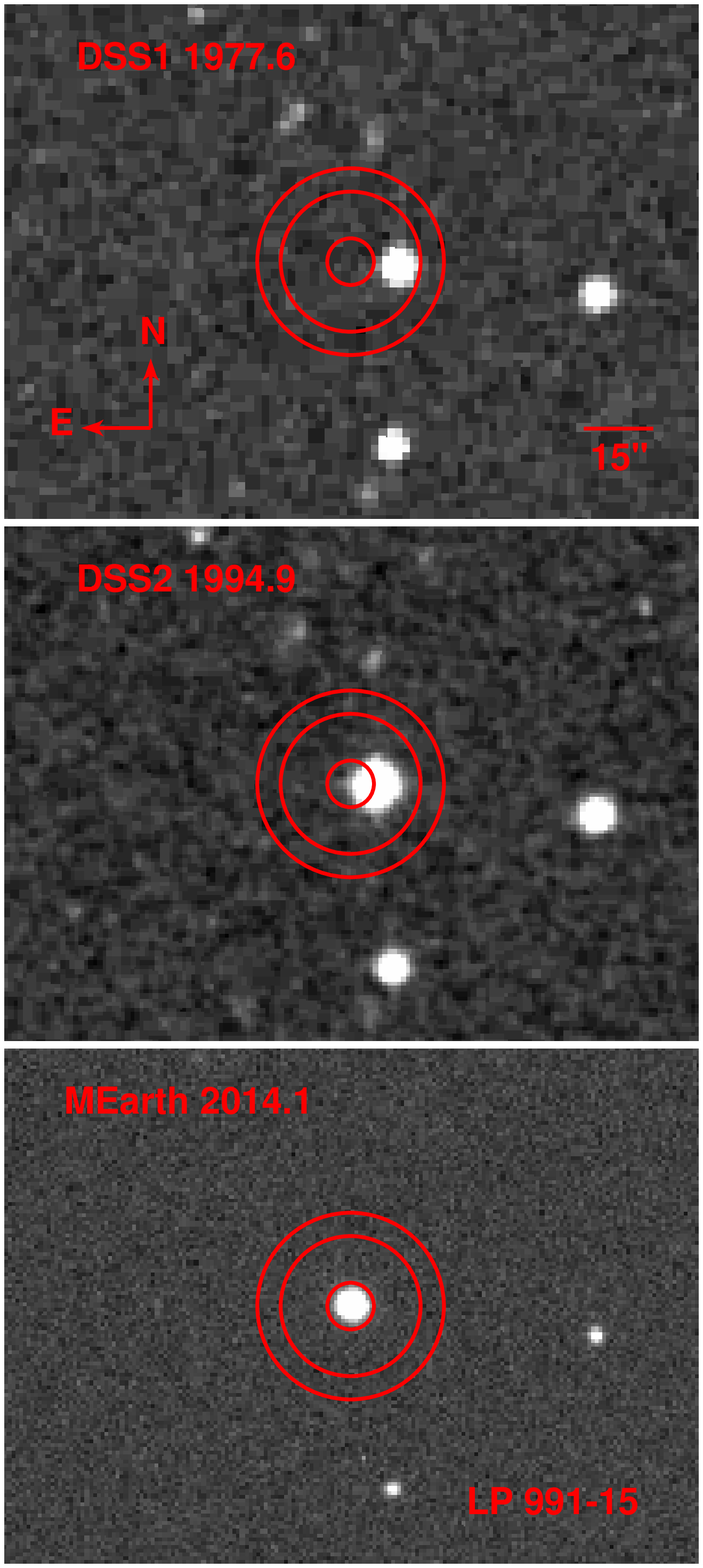}
\caption{Finding charts showing data from the first and second epoch
  Digitized Sky Survey (with the exception of LP~796-24, where the
  first epoch is the SuperCOSMOS scan of the POSS-I ``red'' plate;
  \citealt{2001MNRAS.326.1279H}) and the MEarth master image, in
  time order from top to bottom.  The circles show the approximate
  position and size of the photometric aperture and sky annulus used
  to produce light curves from the MEarth images.  All of the panels
  have the same orientation and scale on sky, the differences in size
  of the aperture and sky annulus are due the use of a different types
  of CCDs for MEarth-North and MEarth-South, which have different
  pixel sizes.}
\label{charts}
\end{figure*}

Finding charts for all four objects, showing the position and size of
the MEarth photometric apertures as well as source proper motion using
previous epochs of imaging, are shown in Figure \ref{charts}.  The
MEarth aperture for LP~107-25 is contaminated by a fainter background
source but apertures for the other targets appear to be clean to the
limiting magnitude of the first epoch plate scans.  LP~261-75 also has
high-contrast imaging observations from \citet{2016ApJ...827..100B}
placing upper limits on the presence of visual companions at very
small angular separations.

The photometric ephemerides, orbital parameters, and rotation periods
for the targets are summarized in Table \ref{ephparams}, where we use 
the final solutions presented in \S \ref{models_sect} and \S
\ref{disc_sect} but combine these parameters into a single table for
convenience.  The quantities given for each system are the epoch of
inferior conjunction $T_0$ and orbital period $P$; the epoch of
superior conjunction $T_{\rm sec}$ for systems with secondary
eclipses; and the photometric rotation period $P_{\rm rot}$ where this
differs from the orbital period.

\begin{deluxetable*}{lrrrr}
\tablecaption{\label{ephparams} Photometric ephemerides, eccentricity and rotation for the four systems.}
\tablecolumns{5}

\tablehead{
\colhead{Parameter}     & \colhead{LP~107-25} & \colhead{LP~261-75} & \colhead{LP~796-24}  & \colhead{LP~991-15}
}

\startdata
$T_0$ (BJD-TDB)         & $2456570.714584$    & $2458159.731511$    & $2457016.818868$    & $2457269.890136$ \\
                        & $\pm 0.000036$      & $\pm 0.000020$      & $\pm 0.000015$      & $\pm 0.000032$ \\
$P$ (days)              & $1.388417440$       & $1.8817205$         & $0.523438589$       & $29.2678016$ \\
                        & $\pm 0.000000074$   & $\pm 0.0000011$     & $\pm 0.000000014$   & $\pm 0.0000081$ \\
\\
$T_{\rm sec}$ (BJD-TDB) & $2456571.407410$    & \ldots              & $2457017.080587$    & \ldots \\
                        & $\pm 0.000069$      & \ldots              & $\pm 0.000015$      & \ldots \\
\\
Eclipse geometry        & total               & total               & grazing             & grazing \\
\\
$P_{\rm rot}$ (days)    & synchronized        & $2.22$              & synchronized        & $34$ \\
\enddata

\tablecomments{We refer the reader to \citet{2011ApJ...727...56I}
  for a discussion of the uncertainties in rotation periods derived
  from MEarth data.}

\end{deluxetable*}

We note that our models have been adjusted for light travel time
across the eclipsing system (in the solar system, this effect is
called the R{\o}mer delay).  This correction is needed for precise
observations of systems with two eclipses to avoid erroneously
inferring small amounts of eccentricity ($e \cos \omega$) due to the
displacement of the secondary eclipse time that results from this
signal being emitted further away from the observer than the primary
eclipse signal in systems where $q < 1$.  The conjunction times $T_0$
and $T_{\rm sec}$ presented in the table are reckoned as if they were
communicated to the observer by a light signal emitted from the
barycenter of the eclipsing system.  They are not eclipse times (as
would be observed) where the eclipse signals originate from the
position of the star closer to the observer.  These corrections depend
on the mass ratio and physical size of the system and are below $1$
minute for all of the systems presented here, but may need to be
applied for precise work.

In several objects the uncertainties on $T_0$ from the Monte Carlo
procedure are very small.  It is likely these have been underestimated
due to correlated noise, which has not been taken into account in this
analysis.  The contribution of systematic error in the shutter timing
also begins to become important at this level, and is not known to
better than approximately $1$ second.

\section{Spectroscopic observations}

For the three systems readily accessible from the Northern hemisphere
(LP~107-25, LP~261-75, and LP~796-24), spectroscopic observations were
gathered using the Tillinghast Reflector Echelle Spectrograph (TRES)
on the 1.5m Tillinghast reflector at Fred Lawrence Whipple Observatory
(FLWO).  Typical exposure times were $3 \times (900-1300)$ s per epoch
using the medium fiber ($R \simeq 44\,000$) and were varied at the
telescope depending on observing conditions.

LP~991-15 is also accessible with TRES, but only at high airmass and
for a limited amount of time per night.  Due to the anticipated
need for a large number of long exposures on this system to measure
the spectroscopic orbit, we opted instead to use the CHIRON instrument
on the SMARTS 1.5m telescope at Cerro Tololo Inter-American
Observatory (CTIO).  Typical exposure times were $3 \times 1200$ s per
epoch in fiber mode ($R \simeq 25\,000$).

These spectrographs are quite similar, so the remainder of the
discussion has been combined and is presented in parallel for both
instruments.

ThAr frames were taken before and after each set of exposures and used
for wavelength calibration.  This is standard operating procedure for
TRES but should be explicitly requested for CHIRON.  Data were reduced
using the standard reduction pipelines provided for both instruments
\citep{2010ApJ...720.1118B,2013PASP..125.1336T}.

The extracted spectra were not blaze corrected to preserve the photon
weighting in the cross-correlations, but the blaze function derived
from the flat fields was retained for later use.  We note that the
extracted spectra are not sky subtracted for either instrument, and in
the case of CHIRON the instrumental background also appears to not be
removed and adds substantially to the counts seen in the extracted
spectra, particularly for long exposures.  The residual instrumental
background is successfully removed by our standard background
subtraction prior to cross-correlation, for which we use quartic
Legendre polynomials.  In order to prevent contamination from sky
emission lines or cosmic ray hits, we reject emission features using
$5 \sigma$ clipping prior to cross-correlation.

Observed template spectra were obtained for use in the
cross-correlation analysis.  For both instruments these were high
signal to noise ratio spectra of Barnard's Star (Gl 699), obtained on
UT 2011 April 15 for TRES and UT 2014 September 18 for CHIRON.
Following our earlier work we adopt a Barycentric
radial velocity of $-110.3 \pm 0.5\ \kms$ for Barnard's Star, where
the stated uncertainty reflects our estimate of the systematic error.
We note that this velocity zero point error propagates to all
``absolute'' Barycentric velocities given in this paper, and in
particular it usually dominates the final uncertainty in the $\gamma$
velocity.  In the tables we have given only the random error from the
Monte Carlo analysis, but note that for most applications this
should be combined with the systematic error in the velocity zero
point.

For cross-correlation analysis to determine radial velocities, we use
a single order of the spectrum close to 7100$\mbox{\AA}$ dominated by
strong molecular features (mostly due to TiO).  For TRES this is the
41st order in the extracted spectrum file (numbering from 1 for the
bluest extracted order), where we use a wavelength range of
7065--7165$\mbox{\AA}$.  This range removes part of the red end of the
order, which is contaminated by telluric absorption features.  For
CHIRON we use the 46th extracted order with a wavelength range of
7040--7120$\mbox{\AA}$.  The signal to noise ratios of the target star
spectra were approximately $10-30$ per pixel in this wavelength range
(the pixel scale is $0.06{\rm \mbox{\AA}/pix}$ for TRES and $0.10{\rm
  \mbox{\AA}/pix}$ for CHIRON) except for LP~796-24 which had a signal
to noise ratio of approximately $5$ per pixel.

\vspace{3ex}  

\section{Spectroscopic analysis}
\label{vrad_sect}

Our initial ``reconnaissance'' procedure for suspected spectroscopic
binaries is as follows.  We aim to obtain exposures at one or both
quadratures (as estimated from the photometric ephemeris, assuming a
circular orbit as necessary in systems with only a single eclipse) in
order to maximize separation of the spectral lines in systems with
composite spectra.  These are analyzed using standard
cross-correlation procedures \citep{1998PASP..110..934K}.  We also use
least-squares deconvolution (LSD; \citealt{1997MNRAS.291..658D}) for
visualization purposes, which gives higher velocity resolution for
very closely separated lines at the cost of being more sensitive to
noise.

All four systems have H$\alpha$ emission, and in cases where this
emission is strong the emission line can have a much higher signal to
noise ratio than the surrounding continuum or the regions used for the
LSD, so this feature is also examined.  It should be noted that these
emission lines have an intrinsically broad line profile with a
non-Gaussian form and originate in the chromosphere so are not
necessarily at precisely the same radial velocity as the photosphere.
Our spectra were extracted using cosmic ray rejection, which is known
to affect strong emission features, and are neither flux calibrated
nor background subtracted, so we do not attempt to provide
quantitative measurements of this emission feature.  The H$\alpha$
region in particular is contaminated by uncorrected sky emission lines
in both instruments, in addition to the large additive instrumental
background in the case of the CHIRON observations.

The appropriate method for extraction of radial velocities depends on
the number of spectroscopic components found.  We now discuss the
systems in order of increasing complexity (number of components).
The radial velocities are given in Table \ref{rv_data}.  We use the
symbols $v_j$ for the radial velocity of star $j$ and $h$ for the
cross-correlation at the best-fitting radial velocity (normalized to
unity) throughout this section.

\begin{deluxetable}{lrrrrr}
\tablecaption{\label{rv_data} Radial velocity data.}
\tablecolumns{6}

\tablehead{
\colhead{BJD-TDB} & \colhead{$v_1$} & \colhead{$v_2$} &
\colhead{$v_3$} & \colhead{$h$} & \colhead{$t_{\rm exp}$}\\ 
\colhead{(days)} & \colhead{(${\rm km/s}$)} & \colhead{(${\rm
    km/s}$)} & \colhead{(${\rm km/s}$)} & & \colhead{(s)}
}

\startdata
\hline
\multicolumn{6}{l}{LP~107-25 ($\alpha = 0.0748$, $\beta = 0.2048$,
  $v_{\rm b1},v_{\rm b2} = 15.5, 6.7\ {\rm km/s}$)}\\
\hline
$2456560.6761$ &$33.524$ &$-124.200$ &$-9.361$ &$0.9435$ &$2700$ \\
$2456573.7180$ &$-40.958$ &$94.424$ &$-14.312$ &$0.8713$ &$2700$ \\
$2456574.7155$ &$22.625$ &$-87.527$ &$-14.641$ &$0.9280$ &$2700$ \\
$2456575.7592$ &$25.449$ &$-90.200$ &$-14.774$ &$0.8971$ &$3600$ \\
$2456576.7053$ &$-43.197$ &$103.919$ &$-15.159$ &$0.9227$ &$3000$ \\
$2456578.6669$ &$35.967$ &$-121.785$ &$-16.004$ &$0.9327$ &$2700$ \\
$2457583.9030$ &$38.591$ &$-119.870$ &$-22.153$ &$0.8683$ &$3000$ \\
\hline
\multicolumn{6}{l}{LP~261-75 ($v_{\rm b1} = 7.57\ {\rm km/s}$)}\\
\hline
$2458082.9421$ &$-25.724$ &$\ldots$ &$\ldots$ &$0.8750$ &$3600$ \\
$2458083.9371$ &$16.321$ &$\ldots$ &$\ldots$ &$0.8922$ &$3600$ \\
$2458107.0107$ &$-2.686$ &$\ldots$ &$\ldots$ &$0.8374$ &$3600$ \\
$2458107.9946$ &$-4.583$ &$\ldots$ &$\ldots$ &$0.8233$ &$3600$ \\
$2458119.0290$ &$-21.240$ &$\ldots$ &$\ldots$ &$0.8719$ &$3900$ \\
$2458119.9922$ &$9.527$ &$\ldots$ &$\ldots$ &$0.8693$ &$3900$ \\
$2458172.8909$ &$-4.056$ &$\ldots$ &$\ldots$ &$0.8448$ &$3600$ \\
$2458187.7043$ &$11.282$ &$\ldots$ &$\ldots$ &$0.8778$ &$3600$ \\
\hline
\multicolumn{6}{l}{LP~991-15 ($\alpha = 0.6510$)}\\
\hline
$2456948.5523$ &$-2.879$ &$22.267$ &$\ldots$ &$0.8910$ &$3600$ \\
$2456951.5435$ &$-16.642$ &$38.432$ &$\ldots$ &$0.8927$ &$3600$ \\
$2456954.5368$ &$-9.545$ &$30.287$ &$\ldots$ &$0.9188$ &$3600$ \\
$2456955.5368$ &$-7.524$ &$27.199$ &$\ldots$ &$0.9146$ &$3600$ \\
$2456957.6848$ &$-2.192$ &$21.981$ &$\ldots$ &$0.8679$ &$3600$ \\
$2456965.5309$ &$14.733$ &$1.205$ &$\ldots$ &$0.9219$ &$3600$ \\
$2456972.6082$ &$32.169$ &$-18.928$ &$\ldots$ &$0.9240$ &$3600$ \\
$2456976.6687$ &$19.769$ &$-4.407$ &$\ldots$ &$0.8909$ &$3600$ \\
$2456978.6783$ &$-13.175$ &$34.477$ &$\ldots$ &$0.8927$ &$3600$ \\
$2456996.5986$ &$18.827$ &$-2.756$ &$\ldots$ &$0.8650$ &$3600$ \\
$2456998.5887$ &$23.689$ &$-8.781$ &$\ldots$ &$0.8838$ &$3600$ \\
$2457002.5769$ &$33.801$ &$-20.652$ &$\ldots$ &$0.9027$ &$3600$ \\
$2457003.6872$ &$35.629$ &$-22.681$ &$\ldots$ &$0.8976$ &$3600$ \\
$2457004.6962$ &$33.558$ &$-20.900$ &$\ldots$ &$0.9022$ &$3600$ \\
$2457005.5698$ &$25.569$ &$-10.817$ &$\ldots$ &$0.8625$ &$3600$ \\
$2457008.6712$ &$-16.636$ &$39.062$ &$\ldots$ &$0.8931$ &$3600$ \\
$2457009.5970$ &$-17.367$ &$39.107$ &$\ldots$ &$0.9097$ &$3600$ \\
$2457010.6070$ &$-15.547$ &$37.708$ &$\ldots$ &$0.9167$ &$3600$ \\
$2457013.5569$ &$-8.348$ &$28.807$ &$\ldots$ &$0.9092$ &$3600$ \\
$2457014.5480$ &$-6.102$ &$25.704$ &$\ldots$ &$0.8468$ &$3600$ \\
$2457015.5512$ &$-3.587$ &$23.314$ &$\ldots$ &$0.8868$ &$3600$ \\
$2457016.5503$ &$-1.238$ &$20.438$ &$\ldots$ &$0.9020$ &$3600$ \\
$2457017.6878$ &$1.135$ &$17.841$ &$\ldots$ &$0.8831$ &$3600$ \\
$2457018.5646$ &$3.278$ &$15.746$ &$\ldots$ &$0.8702$ &$3600$ \\
$2457031.6008$ &$33.469$ &$-20.823$ &$\ldots$ &$0.8781$ &$3600$ \\
$2457032.6102$ &$35.108$ &$-22.568$ &$\ldots$ &$0.8943$ &$3600$ \\
$2457033.5890$ &$35.156$ &$-22.029$ &$\ldots$ &$0.9177$ &$3600$ \\
$2457047.5634$ &$2.372$ &$15.914$ &$\ldots$ &$0.9201$ &$3600$ \\
$2457180.8935$ &$29.556$ &$-15.840$ &$\ldots$ &$0.8678$ &$3600$ \\
$2457182.8806$ &$-5.700$ &$26.214$ &$\ldots$ &$0.8342$ &$3600$ \\
\enddata

\end{deluxetable}

\subsection{Single-lined system (LP~261-75)}

LP~261-75 was found to be single-lined, with some rotational
broadening.  Radial velocities were obtained from $8$ epochs using
standard cross-correlation analysis with rotational broadening applied
to the template spectrum immediately prior to correlation.  The
appropriate amount of rotational broadening\footnote{Where we denote
  the rotational broadening applied to star $j$ as $v_{{\rm b}j}$.  We
  make an explicit distinction between {\it assumed} or {\it adopted}
  broadening in the correlation analysis using this symbol and {\it
    true} rotational broadening $v_{\rm rot} \sin i$, which is not
  necessarily the same, as discussed in the text.} $v_{\rm b1}$ was
determined by searching for maximum peak correlation as a function of
$v_{\rm b1}$ for each epoch individually, adopting the mean value of
$v_{\rm b1} = 7.57 \pm 0.10\ \kms$ (where the stated uncertainty is
the empirical standard error in the mean calculated from the sample of
$8$ measurements) for the final analysis of all the epochs.

We note that this value of $v_{\rm b1}$ is only barely above the
spectral resolution, and could be influenced by other sources of
broadening than rotation.  The treatment of rotational broadening in
the analysis is based on sampling the spectrum in $\log \lambda$ at
approximately $1/32$ pixel and neglects any rotational broadening
which might be present in the template spectrum, so it is only
approximate for small $v_{\rm b1}$ and could potentially also impact
the results.  This should therefore not be treated as a measurement of
the rotational broadening.

Radial velocity uncertainties were derived from the scatter in the
residuals during fitting and were approximately $0.14\ \kms$ for this
system.

\subsection{Double-lined system (LP~991-15)}

LP~991-15 was observed after only a single eclipse had been detected,
so the orbital period was unknown and it was not possible to arrange
to take data at the optimum orbital phase.  After obtaining several
epochs with insufficient velocity separation between the components we
eventually determined this object to be double-lined, with negligible
rotational broadening, and obtained a total of $38$ epochs.  Radial
velocities were derived using TODCOR \citep{1994ApJ...420..806Z},
following the procedures in \citet{2011ApJ...742..123I}.  Eight epochs
with velocity separation $|v_1-v_2| < 10\ \kms$ were
discarded, leaving $30$ epochs for the final analysis.  The
spectroscopic light ratio $\alpha$ was derived by searching for the
maximum sum of the squares of the peak correlation over all the
remaining epochs, and gave $\alpha = 0.6510$.  The radial velocity
uncertainties derived during fitting were approximately $0.16\ \kms$
for the primary and $0.31\ \kms$ for the secondary.

\subsection{Triple-lined systems (LP~107-25 and LP~796-24)}

LP~107-25 and LP~796-24 were both found to be triple-lined with
distant, slowly rotating companions, and an inner, rapidly rotating
eclipsing binary pair.  Throughout the discussion of these objects, we
refer to the eclipsing binary pair as the ``primary'' and
``secondary'', and the third star as the ``tertiary'', with respective
indices $1$, $2$ and $3$, even though in the case of LP~796-24 it is
possible the star we refer to as the ``tertiary'' is the most massive.

Radial velocities for these triple-lined systems were derived using
TRICOR \citep{1995ApJ...452..863Z}, which is the extension of the
TODCOR method to three dimensions.  This requires three template
spectra and two light ratio parameters $\alpha$ and $\beta$.  In the
present case all three templates were the same spectrum of Barnard's
Star but were allowed to have different amounts of rotational
broadening.  As for the analysis of LP~991-15, two epochs for
LP~107-25 with velocity separation $|v_1-v_2| < 10\ \kms$ were
discarded.

In both objects star $3$ was found to have negligible rotational
broadening, so none was applied to the template, and stars $1$ and $2$
were found to be rotating synchronously within the uncertainties, so
the $v_{\rm b}$ values for these stars were fixed to the values
calculated from the models presented in the following sections,
leaving only the two parameters $\alpha$ and $\beta$ to be determined
using TRICOR.

For LP~107-25, the velocity uncertainties derived during fitting were
approximately $1.6\ \kms$ for the primary and tertiary, and
$2.5\ \kms$ for the fainter secondary.

For LP~796-24, we only obtained $3$ usable epochs with low signal to
noise ratios, which yielded unusually low peak correlation.  The
velocities and light ratios do not seem to be reliably determined.
We therefore do not present them in the table, and due to the lack of
spectroscopic information needed for a full solution of this system,
we only present the ephemeris and show the MEarth light curve in this
paper.

\section{Models}
\label{models_sect}

We use a simplified version of the procedure described in
\citet{2011ApJ...742..123I} to model the light curves and radial
velocities simultaneously for the multiple-lined systems, which
is based on the Nelson-Davis-Etzel model
\citep{1972ApJ...174..617N,1981psbs.conf..111E,1981AJ.....86..102P}
and its descendant JKTEBOP \citep{2004MNRAS.351.1277S,2004MNRAS.355..986S}.
Since the \citet{2011ApJ...742..123I} publication, the light curve
generator was rewritten\footnote{This software is available online at
  the following URL: \url{https://github.com/mdwarfgeek/eb}}, and now
uses the analytic method of \citet{2002ApJ...580L.171M} to perform the
eclipse calculations.  This model is physically equivalent but
avoids the trade-off between performance and accuracy inherent in the
original implementation (due to use of numerical integration).

Table \ref{model_pars} summarizes the parameters in the models and
their symbols used in the text and tables, and Table
\ref{adopted_pars} gives the values adopted for each system (excluding
LP~796-24, where we did not undertake a full solution).  We refer the
reader to \citet{2011ApJ...742..123I} for a discussion of the choice
of priors for these parameters, which have been adopted here except as
noted in the text.  The values of the modified Jeffreys prior
parameter $K_a$ used in the radial velocity analysis were set to
$10\%$ of the $s_j$ parameters for the radial velocities.

\begin{deluxetable*}{ll}
\tablecaption{\label{model_pars} Definition of parameters used in the models.}
\tablecolumns{2}

\tablehead{
\colhead{Parameter} & \colhead{Description}
}

\startdata
$J$                 & Central surface brightness ratio (secondary / primary) in MEarth. \\
$(R_1+R_2)/a$       & Sum of the component radii $R_1$ and $R_2$ divided by semimajor axis $a$. \\
$R_2/R_1$           & Radius ratio. \\
$\cos i$            & Cosine of orbital inclination $i$. \\
$e \cos \omega$     & Eccentricity $e$ multiplied by cosine of argument of periastron $\omega$. \\
$e \sin \omega$     & Eccentricity multiplied by sine of argument of periastron. \\
\\
$u_1$               & Linear limb darkening coefficient for primary. \\
$u'_1$              & Quadratic limb darkening coefficient for primary. \\
$u_2$               & Linear limb darkening coefficient for secondary. \\
$u'_2$              & Quadratic limb darkening coefficient for secondary. \\
$(y\beta)_1$        & Product of gravity darkening coefficient and exponent for primary. \\
$(y\beta)_2$        & Product of gravity darkening coefficient and exponent for secondary. \\
$A_1$               & Albedo of primary. \\
$A_2$               & Albedo of secondary. \\
\\
$L_2/L_1$           & Light ratio (secondary/primary). \\
$L_3/L_{\rm tot}$   & Third light divided by total system light. \\
\\
$F_1$               & Rotation parameter of primary (orbital period / rotation period). \\
$a_{11}$            & Primary out-of-eclipse sine coefficient for fundamental. \\
$b_{11}$            & Primary out-of-eclipse cosine coefficient for fundamental. \\
$a_{12}$            & Primary out-of-eclipse sine coefficient for 2nd harmonic. \\
$b_{12}$            & Primary out-of-eclipse cosine coefficient for 2nd harmonic. \\
\\
$K_1$               & Radial velocity semiamplitude of primary (for SB1s). \\
\\
$q$                 & Mass ratio, $M_2 / M_1$. \\
$K_1+K_2$           & Sum of radial velocity semiamplitudes (for SB2s). \\
\\
$\gamma$            & Systemic radial velocity. \\
\\
$P$                 & Orbital period. \\
$T_0$               & Epoch of inferior conjunction. \\
\\
$z_j$               & Magnitude zero point for light curve segment $j$ (see \S \ref{lc_nuisance_sect}).\\
$s_j$               & For light curves, error scaling factor for light curve $j$.\\
                    & For radial velocities, this parameter sets the adopted radial velocity \\
                    & uncertainty according to $\sigma_{ij} = s_j / h_i$ for data point $i$ of star $j$, \\
                    & where $h_i$ is the peak normalized cross-correlation from Table \ref{rv_data}. \\
$C$                 & Common mode coefficient (described in \S \ref{lc_nuisance_sect}). \\
\enddata

\tablecomments{This table updates Table 5 from
  \citet{2011ApJ...742..123I}.  We have changed some of the symbols
  and amended the descriptions following changes to the software, but
  the parameter set is reproduced here in full for convenience.}

\end{deluxetable*}

\begin{deluxetable}{lrrr}
\tablecaption{\label{adopted_pars} Adopted parameters for each system.}
\tablecolumns{4}

\tablehead{
\colhead{Parameter}                            & \colhead{LP~107-25}  & \colhead{LP~261-75}  & \colhead{LP~991-15}
}

\startdata
$J$                  & varied             & $0$             & varied \\
$(R_1+R_2)/a$        & varied             & varied          & varied \\
$R_2/R_1$            & varied             & varied          & $0.871 \pm 0.050$ \\
$\cos i$             & varied             & varied          & varied \\
$e \cos \omega$      & varied             & $0$             & varied \\
$e \sin \omega$      & varied             & $0$             & varied \\
\\
$T_{{\rm eff}1}$ (K) & $3500$             & $3100$          & $3150$ \\
$T_{{\rm eff}2}$ (K) & $3050$             & \ldots          & $3050$ \\
\\
$u_1$                & $0.1857$           & $0.2352$        & $0.2269$ \\
$u'_1$               & $0.3205$           & $0.4008$        & $0.3932$ \\
$u_2$                & $0.2856$           & $0$             & $0.2856$ \\
$u'_2$               & $0.4549$           & $0$             & $0.4549$ \\
$(y\beta)_1$         & $0.2478$           & $0$             & $0.2978$ \\
$(y\beta)_2$         & $0.3203$           & $0$             & $0.3203$ \\
$A_1$                & $0.4$              & $0$             & $0.4$ \\
$A_2$                & $0.4$              & $0$             & $0.4$ \\
\\
$L_2/L_1$            & $0.075 \pm 0.050$  & $0$             & $0.651 \pm 0.050$ \\
$L_3/L_{\rm tot}$    & $0.160 \pm 0.050$  & $0$             & $0$ \\
\\
$F_1$                & $1.0$              & varied          & varied \\
$a_{11}$             & varied             & varied          & varied \\
$b_{11}$             & varied             & varied          & varied \\
$a_{12}$             & $0$                & varied          & $0$ \\
$b_{12}$             & $0$                & varied          & $0$ \\
\\
$K_1$                & \ldots             & varied          & \ldots \\
\\
$q$                  & varied             & \ldots          & varied \\
$K_1+K_2$            & varied             & \ldots          & varied \\
\\
$\gamma$             & varied             & varied          & varied \\
\\
$P$                  & varied             & varied          & varied \\
$T_0$                & varied             & varied          & varied \\
\enddata

\tablecomments{In this table, values without uncertainties indicate
  the parameter was fixed at this value, and values with uncertainties
  indicate the parameter was varied subject to a Gaussian prior with
  the mean and standard deviation given.  The effective temperatures
  given are the values assumed when interpolating limb darkening and
  gravity darkening coefficients.}

\end{deluxetable}

Similar to our analysis of other single-lined systems
(e.g., \citealt{2010ApJ...718.1353I} and the MEarth transiting planets)
a light curve model with a completely dark secondary was used for
LP~261-75, with the photometric ephemeris imposed on the radial
velocity solution using priors for simplicity given that the solution
of such systems is essentially separable.  This used the same
underlying model as for the multiple-lined systems, but neglecting
ellipsoidal variation and light travel time (due to lack of knowledge
of the radial velocity semiamplitude of the secondary), the reflection
effect, and gravity darkening.  The radial velocity model fitting for
this single-lined system was done using the same implementation as
\citet{2018AJ....155..125W} with priors as described there.

In the following subsections we highlight several points of note
specific to the present analysis.

\subsection{Limb darkening and gravity darkening}

Our photometry was all gathered using MEarth, which was changed to a
non-standard filter bandpass during the 2011 summer shutdown and used
for all data analyzed here, so limb darkening coefficients now require
more careful treatment.  In addition, as discussed by
\citet{2017ApJ...836..177T}, gravity darkening coefficients are
needed.

To allow the use of limb darkening and gravity darkening coefficients
from standard tabulations, transformation equations were derived from
the Sloan Digital Sky Survey $i$ and $z$ filters to MEarth.  These
calculations were performed with the Limb Darkening Toolkit (LDTK;
\citealt{2015MNRAS.453.3821P}) using PHOENIX model atmospheres from
\citet{2013A&A...553A...6H} and the MEarth transmission function from
\citet{2016ApJ...818..153D}.  We use the quadratic limb darkening law
throughout the calculations.

Over the appropriate range of spectral type needed in this work, the
following equation was found to be sufficient:
\begin{equation}
u({\rm MEarth}) = \left[2 u(i) + 3 u(z)\right] / 5
\label{ld_eq}
\end{equation}
where $u({\rm passband})$ refers to the limb darkening coefficient in
each passband, and the transformation equation is the same for the
linear and quadratic coefficients, which we denote as $u$ and $u'$ in
this work due to use of the usual symbols for other purposes.  We
further assume that the same transformation applies to the gravity
darkening coefficients.

Limb darkening coefficients were adopted from the tables of
\citet{2012A&A...546A..14C}, using fixed values of $\log g = 5.0$, and
the least-squares method, and interpolated in effective temperature
$T_{\rm eff}$.  Gravity darkening coefficients were adopted from
\citet{2011A&A...529A..75C} using the same value of $\log g$ and
PHOENIX atmosphere models but interpolated in the native $\log T_{\rm
  eff}$ abscissa used in the table.  These tables give the quantity
$y(\lambda) \beta$ in the notation of \citet{2017ApJ...836..177T}
needed as input to the light curve model directly (we note that the
definition of the symbol $y$ used in \citealt{2011A&A...529A..75C}
differs from the definition we and \citealt{2017ApJ...836..177T}
use).

Effective temperatures are not well-constrained for any of our
targets, so we use the tables of \citet{2013ApJS..208....9P} for main
sequence stars\footnote{From \url{http://www.pas.rochester.edu/~emamajek/EEM\_dwarf\_UBVIJHK\_colors\_Teff.txt} version 2017.10.19.}
to estimate $T_{\rm eff}$ with linear interpolation.  For the
multiple-lined systems LP~107-25 and LP~991-15 we interpolate using
the measured component masses.  For the single-lined system LP~261-75
we use the spectral type given in Table \ref{photparams}.  The adopted
values for all four objects are given in Table \ref{adopted_pars}.
Precision in excess of $50\ {\rm K}$ is not warranted, especially
given the intrinsic scatter and propensity for systematic error in
effective temperature scales for M-dwarfs, so the $T_{\rm eff}$ values
were rounded to the nearest $50\ {\rm K}$.

\subsection{Third light}

Two of our systems (LP~107-25 and LP~796-24) have a large amount of
third light, which dilutes the measured eclipse depths and must be
accounted for in the light curve solutions.  The determination of this
quantity from otherwise unconstrained light curve models is a
notoriously degenerate problem (e.g., \citealt{1972ApJ...174..617N})
and it is likely our ground-based light curves lack the precision
required to attempt this.

In order to make progress, we use the spectroscopic measurements.  The
quantity required as input to the model is the third light divided by
total light in the observed photometric bandpass, which we denote as
$L_3 / L_{\rm tot}$ where $L_j$ is the light of the $j$th star, in
arbitrary units, and $L_{\rm tot} = L_1 + L_2 + L_3$.  The TRICOR
solutions provide two light ratios $\alpha \simeq L_2 / L_1$ and
$\beta \simeq L_3 / L_1$ where the approximations serve to emphasize
the assumption that the effective bandpass of the spectroscopic
measurement matches that of the photometry.  This is not the case, so
the resulting value of the third light parameter inherits an
uncertainty from this procedure which we discuss in more detail in \S
\ref{spec_vs_phot_sect}.  We apply this constraint using a Gaussian
prior, adopting a standard deviation of $0.05$.

The LP~261-75 and LP~991-15 systems do not show any evidence of third
light in the spectroscopy or in imaging observations, so we adopt a
fixed value of $L_3 = 0$.

\subsection{Light ratios}

As discussed in \citet{2011ApJ...742..123I}, in most grazing
configurations it is also necessary to impose a spectroscopic
constraint on the light ratio $L_2 / L_1$ using the TODCOR (or TRICOR)
$\alpha$ parameter when working with ground-based light curves.

In the case of LP~261-75 there is no evidence for light from the
secondary in the light curves or spectra so this was fixed to zero,
which is equivalent to fixing $J = 0$.  The eclipse is total so the
light curves are then sufficient to constrain all three geometric
parameters $R_2/R_1$, $\cos i$, and $(R_1+R_2)/a$.

In LP~107-25 and LP~991-15, we apply the spectroscopic constraint
using a Gaussian prior with a mean given by the spectroscopic value
and a standard deviation of $0.05$.  It serves different roles in each
object due to their configurations.  In LP~107-25, due to the totally
eclipsing geometry this constraint merely serves to supplement the
information provided by the third light constraint.  For LP~991-15, the
spectroscopic constraint supplies essentially all of the information
on the combination of $J$ and $R_2/R_1$ in this system due to the
grazing geometry combined with the lack of secondary eclipses.

\subsection{Radius ratio}

In LP~991-15, we need an additional constraint due to the lack of
secondary eclipses, which would usually serve to constrain the
parameter $J$ (this is essentially determined by the relative depths
of the primary and secondary eclipses and the adopted limb darkening
law).

To make progress in this system, we use the empirical mass-radius
relation of \citet{2006ApJ...651.1155B} to calculate the radius ratio
from the measured component masses, and impose this as a Gaussian
prior on $R_2/R_1$.  We adopt a standard deviation of $0.05$ on this
radius ratio, based on the observed dispersion in the mass-radius
relation.  It is important to note that the results depend on the
adopted mass-radius relation, the implications of which are discussed
in \S \ref{lp991_15_sect}.

\subsection{Spots}

All four systems show evidence of spots.  The solutions in this work
are preliminary in nature, so we have not undertaken a detailed
treatment of the effect of these spots on the derived parameters.
Instead, we simply adopt a standard ``non-eclipsed spots'' model
in the terminology of \citet{2011ApJ...742..123I}, placing the spots
on the primary star.  In the case of LP~107-25 and LP~261-75 it is
unlikely the spots responsible for the modulations could be on any
other component due to the large light ratios (and for LP~107-25 and
LP~796-24 the synchronous rotation periods argue against the origin of
the modulations being the tertiary, which has no rotational broadening
in the spectroscopy for either object), but we have not attempted to
allow for the effect of spot latitude (relative to the eclipse chord)
by varying the ``fraction of eclipsed spots'' parameter.

We caution that a more detailed treatment of the effect of spots would
be necessary in any possible future use of these systems to derive
precise estimates of light curve parameters and stellar radii.

\subsection{Light curve nuisance parameters}
\label{lc_nuisance_sect}

Light curves from MEarth require correction for magnitude zero point
offsets (predominantly thought to be caused by flat fielding error)
when the target crosses the meridian and also when the instrument is
removed from the telescope for servicing.  In the light curves each
place where a new magnitude zero point is needed is given a unique
integer ``segment number'', and we use these to apply the correction,
adding a new free parameter $z_j$ to the model for the magnitude zero
point in each ``segment'' $j$ appearing in each light curve.

We also inflate the observational uncertainties by adding error
scaling parameters $s_l$ for each light curve $l$ (in the electronic
table provided with the manuscript, these are identified by different
``dataset names'' in the second column).  In the case of LP~261-75,
the datasets observed using different telescopes on the same night
were combined for this purpose so we use one $s$ parameter per night
rather than the usual practice which would have assigned separate
parameters to each telescope.  LP~991-15 was observed in a similar
fashion on two nights but these were found to benefit from being left
with separate $s_l$ coefficients due to tracking problems on specific
telescopes.

Variations in atmospheric water vapor also introduce systematic
variations in the MEarth photometry.  This effect has been discussed
in detail elsewhere
(e.g., \citealt{2011ApJ...727...56I,2016ApJ...821...93N}).  The
photometric corrections are derived by averaging the light curves of
all M-dwarfs observed at a given time from telescopes at the same site
to produce a ``common mode'' light curve.  In order to average as many
target stars as possible this must be done in bins of roughly the
standard observational cadence, which are $0.02$ days.  We find this
sampling is too coarse to reliably correct high-cadence followup light
curves such as used during eclipse windows in the present work, and in
some cases such as LP~261-75 multiple telescopes were used to observe
the same target, which adversely impacts determination of the common
mode itself.  We therefore apply this correction only to long-term
light curves with out-of-eclipse parts of the time series and not the
individual eclipses observed for followup.  We use the symbol $C$ for
the common mode coefficient, which should be the same for all
telescopes observing the same target, so only one coefficient is
needed.

\subsection{Method of solution and uncertainties}

Non-linear least-squares model fitting (using MPFIT;
\citealt{2009ASPC..411..251M}; or {\tt leastsq} from {\tt
  scipy.optimize} in the case of LP~261-75) was performed to
initialize the parameters and covariance matrices prior to the final
Monte Carlo simulations.  These fits used iterative $5 \sigma$ outlier
rejection for the light curves, and the resulting clipped light curves
were the ones used in the Monte Carlo simulations.  These outliers are
predominantly in the out-of-eclipse monitoring portions and usually
correspond to bad images (e.g., pointing errors, tracking problems, or
clouds) and occasional stellar flares.  The appropriate value of
$\sigma$ was calculated using a robust median absolute deviation (MAD)
estimator, scaled to the Gaussian equivalent rms
(e.g., \citealt{1983ured.book.....H}).

Parameters and uncertainties were estimated using Monte Carlo
simulations.  For the multiple-lined systems, we used the same
Adaptive Metropolis method used in \citet{2011ApJ...742..123I},
with chains run for $2 \times 10^6$ steps, discarding the first $50\%$
of these to allow the chain to ``burn in'' before being used for
parameter estimation.  For LP~261-75 we used the {\tt emcee} package
\citep{2013PASP..125..306F} with $250$ walkers each run for a burn in
of $10^4$ steps, followed by $2 \times 10^4$ steps used for parameter
estimation (resulting in a total of $5 \times 10^6$ samples from the
posterior probability density function).  In this paper we report the
median as the central value and the $68.3$ percentile of the absolute
deviation of the posterior samples about the median as the
uncertainty.  This change was made compared to our previous work to
produce symmetric uncertainties and thereby simplify interpretation of
the results.

\vspace{1ex}  

\section{Discussion and orbital solutions}
\label{disc_sect}

In this section, we first discuss a difficulty common to several of
the objects, and then present solutions and discussion for each of the
four objects individually.  We give the jump parameters used in the
Monte Carlo simulations and any derived parameters such as masses
which are determined robustly.  Derived parameters which are not
well-determined are not presented, and while we have given sufficient
information to calculate these we caution against doing so given their
uncertainties.  Consequently we also do not present comparisons to
theoretical models, which would be premature given the preliminary
nature of the solutions.

\subsection{Spectroscopic vs photometric light ratios}
\label{spec_vs_phot_sect}

An important source of uncertainty, and potentially also systematic
error, in the models of the multiple-lined systems presented here
(LP~107-25 and LP~991-15) results from the need to compute light
ratios appropriate for models of the MEarth photometry using
quantities measured spectroscopically with the TODCOR or TRICOR
$\alpha$ and $\beta$ parameters.

These problems arise if the components of the multiple system are not
spectroscopically identical.  In such cases the $\alpha$ or $\beta$
parameters then depend not only on the appropriate light ratio, but
also on the degree to which the depths of the absorption lines
resemble the ones in the template (in this case, Barnard's Star).  The
addition of rotational broadening and the need to correct for
differences between the spectroscopic and photometric bandpasses
present further complications.

We conducted simulations using PHOENIX model atmospheres from
\citet{2013A&A...553A...6H} to estimate the appropriate transformation
between the measured spectroscopic ratios and light ratios in the
MEarth bandpass.  We find that these appear to depend strongly on
metallicity in addition to effective temperature (we speculate that
this may be due to the use of molecular absorption features,
predominantly due to TiO, when computing the spectroscopic ratios).
Given that neither of these quantities are well determined for our
targets, we have not attempted to apply any corrections for the
present analysis.  Instead, we simply use the spectroscopic values
without correction and adopt an uncertainty of $\pm 0.05$, which
we find to be a reasonable approximation to the error introduced by
not applying the corrections over realistic ranges in effective
temperature and metallicity.

This strongly downweights the small value of $\alpha$ for LP~107-25,
where we find the appropriate correction factor is ill-determined due
to the large difference in stellar type between the primary and the
secondary.  In this system, the third light parameter appropriate for
the light curve models is better determined by virtue of a smaller
spectral mismatch between the tertiary and primary so is given more
weight in the solution.  The alternative assumption of a constant
relative error in the light ratio would assign too much weight to
ratios far from unity.

We caution that this procedure is still somewhat arbitrary and has an
impact on the resulting uncertainties in the physical parameters of
interest, which is clearly undesirable.  At present, it is difficult
to make further progress using the same method until effective
temperatures and metallicities can be constrained observationally for
our targets, and even with this information the results would still
depend on the model atmospheres.  A possible solution to the latter
problem would be to use multiple systems which are resolved both
visually and spectroscopically to calibrate empirical relations
between photometric and spectroscopic light ratios, but there are not
currently enough examples of such objects known to attempt to derive
these relations.

A potentially superior approach for the two triple systems with
distant tertiaries (LP~107-25 and LP~796-24) may be to attempt to
resolve the tertiary from the inner binary using imaging observations
in order to measure $L_3/(L_1+L_2)$ directly.  Such observations are
not currently available, and these systems are distant, resulting in
small expected angular separations, but this may be a fruitful avenue
for future work.  Light curves obtained in the same bandpass as the
imaging observations could then be analyzed by imposing the measured
light ratios directly, without incurring transformation uncertainties.

\subsection{LP~107-25}

\begin{figure*}
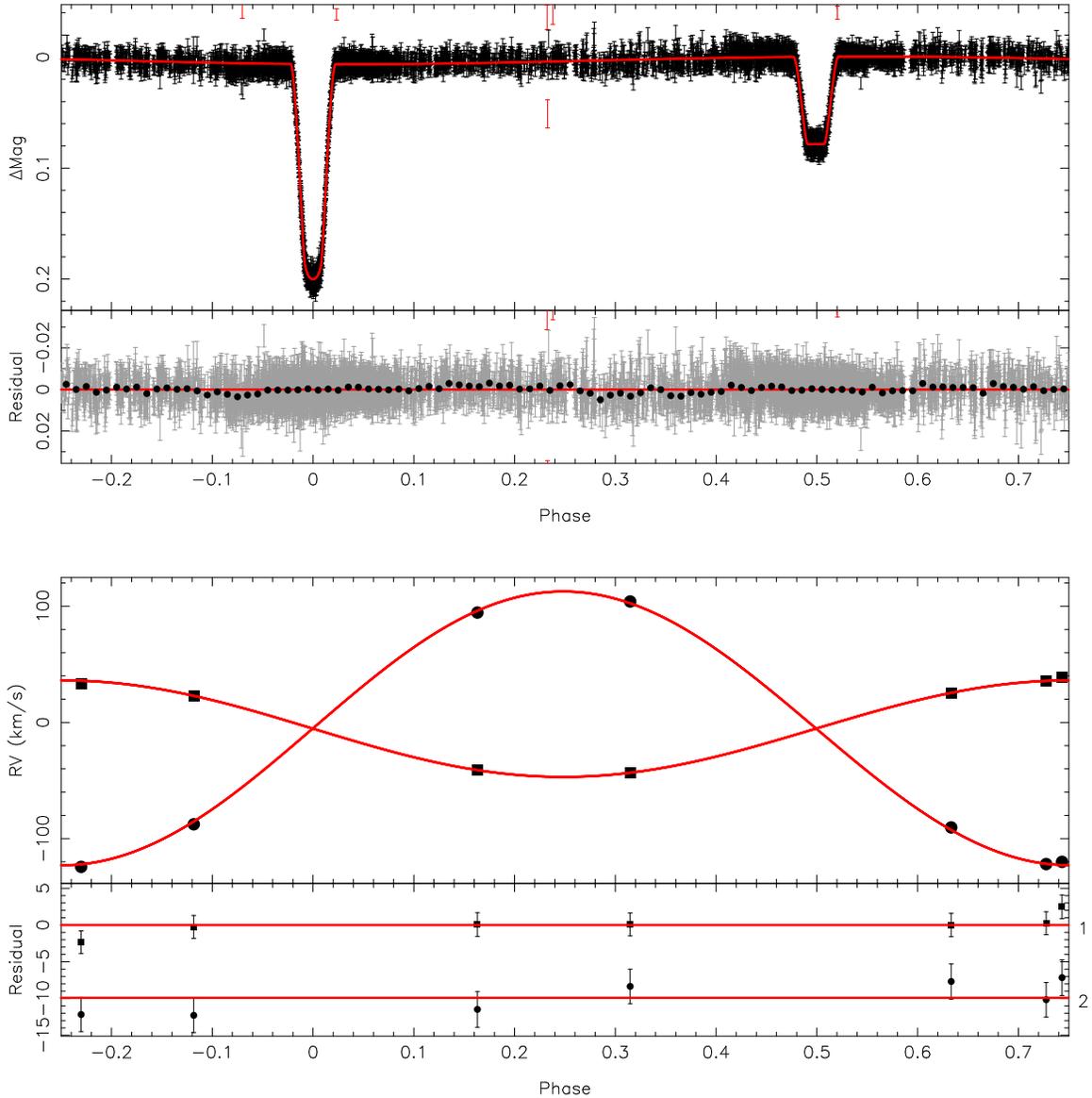

\centering
\includegraphics[angle=270,width=6in]{f2a.eps}

\vspace{4ex}

\includegraphics[angle=270,width=6in]{f2b.eps}

\vspace{2ex}

\caption{Top panels: phase-folded light curve for LP~107-25.  Bottom
  panels: radial velocity curve.  In each case there are two
  sub-panels, with the upper showing the data with the best fit
  overlaid, and the lower showing the residuals from the best fit.
  For the light curves, the magnitude zero point offsets and common
  mode have been removed to flatten the instrumental baseline, leaving
  only suspected astrophysical variations.  Outliers rejected from the
  solution are shown in red.  When plotting the residuals, the
  original data are shown in grey, and black points show the same data
  binned in phase into 100 equal size bins.  For the radial
  velocities, square symbols show the velocities for the primary and
  filled circles for the secondary.  The residuals for the two
  components are offset vertically for clarity.}
\label{lp107_25_modelfit}
\end{figure*}

Figure \ref{lp107_25_modelfit} shows the light curve and radial
velocities for this system, and Table \ref{lp107_25_params} gives our
preliminary orbital solution.  There is a small displacement in the
secondary eclipse timing, where the eclipse appears approximately two
minutes early compared to the prediction for a circular orbit.  We
have allowed non-zero eccentricity in the solution to account for
this, which appears predominantly as a non-zero value of $e \cos
\omega$.  The value of $e \sin \omega$ is consistent with zero within
reasonable uncertainties, especially when accounting for the tendency
of solutions with limited numbers of radial velocities to overfit this
quantity.  Consequently the argument of periastron $\omega$ is
ill-determined and we do not provide individual values for $e$ and
$\omega$ in the table.

\begin{deluxetable}{lr}
\tablecaption{\label{lp107_25_params} Parameters and uncertainties for LP~107-25.}
\tablecolumns{2}

\tablehead{
\colhead{Parameter}     & \colhead{Value}
}

\startdata
\hline
\multicolumn{2}{l}{Jump parameters}\\
\hline
$J$                                  & $0.5035 \pm 0.0029$ \\
$(R_1+R_2)/a$                        & $0.13931 \pm 0.00077$ \\
$R_2/R_1$                            & $0.431 \pm 0.012$ \\
$\cos i$                             & $0.0202 \pm 0.0062$ \\
$e \cos \omega$                      & $-0.001564 \pm 0.000081$ \\
$e \sin \omega$                      & $0.0040 \pm 0.0021$ \\
$q$                                  & $0.3529 \pm 0.0075$ \\
$a_{11}$                             & $-0.00111 \pm 0.00016$ \\
$b_{11}$                             & $-0.00348 \pm 0.00023$ \\
$(K_1+K_2)$ (km/s)                   & $159.3 \pm 1.3$ \\
$\gamma$ (km/s)                      & $-5.29 \pm 0.54$ \\
$C$                                  & $0.557 \pm 0.021$ \\
$s_{1}$ (km/s)                       & $1.50 \pm 0.45$ \\
$s_{2}$ (km/s)                       & $2.26 \pm 0.66$ \\
\hline
\multicolumn{2}{l}{Derived parameters}\\
\hline
$i$ (deg)                            & $88.84 \pm 0.36$ \\
$M_1$ ($\msol$)                      & $0.430 \pm 0.010$ \\
$M_2$ ($\msol$)                      & $0.1518 \pm 0.0046$ \\
$(R_1+R_2)$ ($\rsol$)                & $0.6092 \pm 0.0061$ \\
$R_1$ ($\rsol$)                      & $0.4256 \pm 0.0065$ \\
$R_2$ ($\rsol$)                      & $0.1836 \pm 0.0031$ \\
\enddata

\end{deluxetable}

A background star is seen overlapping the MEarth photometric aperture
in Figure \ref{charts}.  This is not the star responsible for the
third light, and is not included in the TRES fiber, which has a
smaller diameter than the MEarth photometric aperture.  This
contaminating star is $4.72$ magnitudes fainter than the target in the
GAIA $G_{\rm RP}$ passband, and is bluer than the target in the
$G_{\rm BP}-G_{\rm RP}$ color, so the contribution to the MEarth
photometry should be negligible compared to the uncertainty inherited
from the TRICOR-derived value of $\beta / (1 + \alpha + \beta)$ for
the brighter star that is the source of the third light in the
spectroscopy.

We regard our solution as preliminary in both the masses, due to the
limited number and quality of the radial velocities, and the
possibility of a long-term trend due to the outer orbit; and in the
radii due to the large amount of third light contamination.  It is
likely the uncertainties are underestimated as a result of these
issues, and due to neglecting correlated noise in the analysis.
Comparison of these quantities to the predictions of theoretical
models would therefore be premature, and we have not undertaken such
an analysis at present.

\subsection{LP~261-75}
\label{lp261_75_sect}

\begin{figure*}
\centering
\includegraphics[width=3.5in]{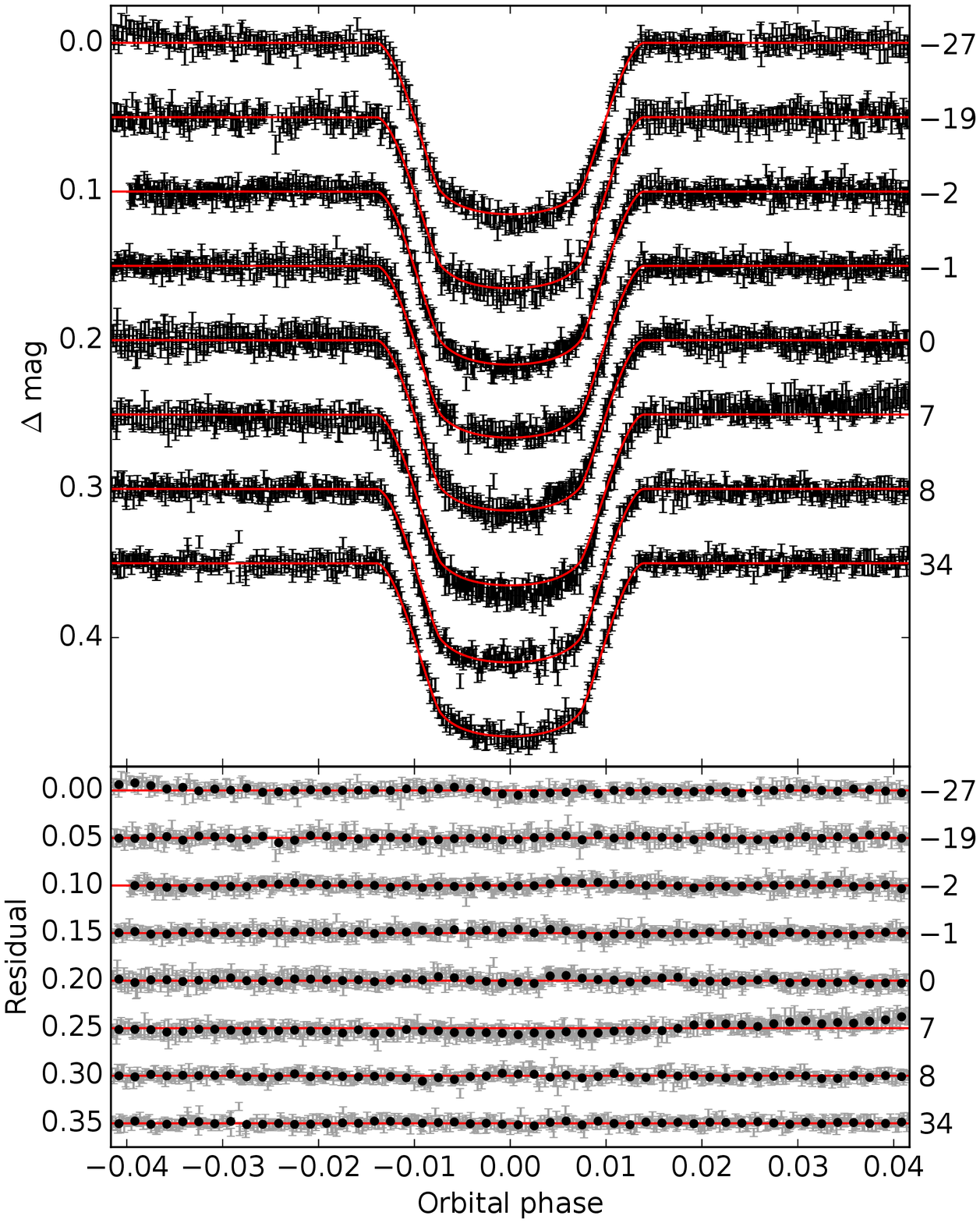}
\includegraphics[width=3.5in]{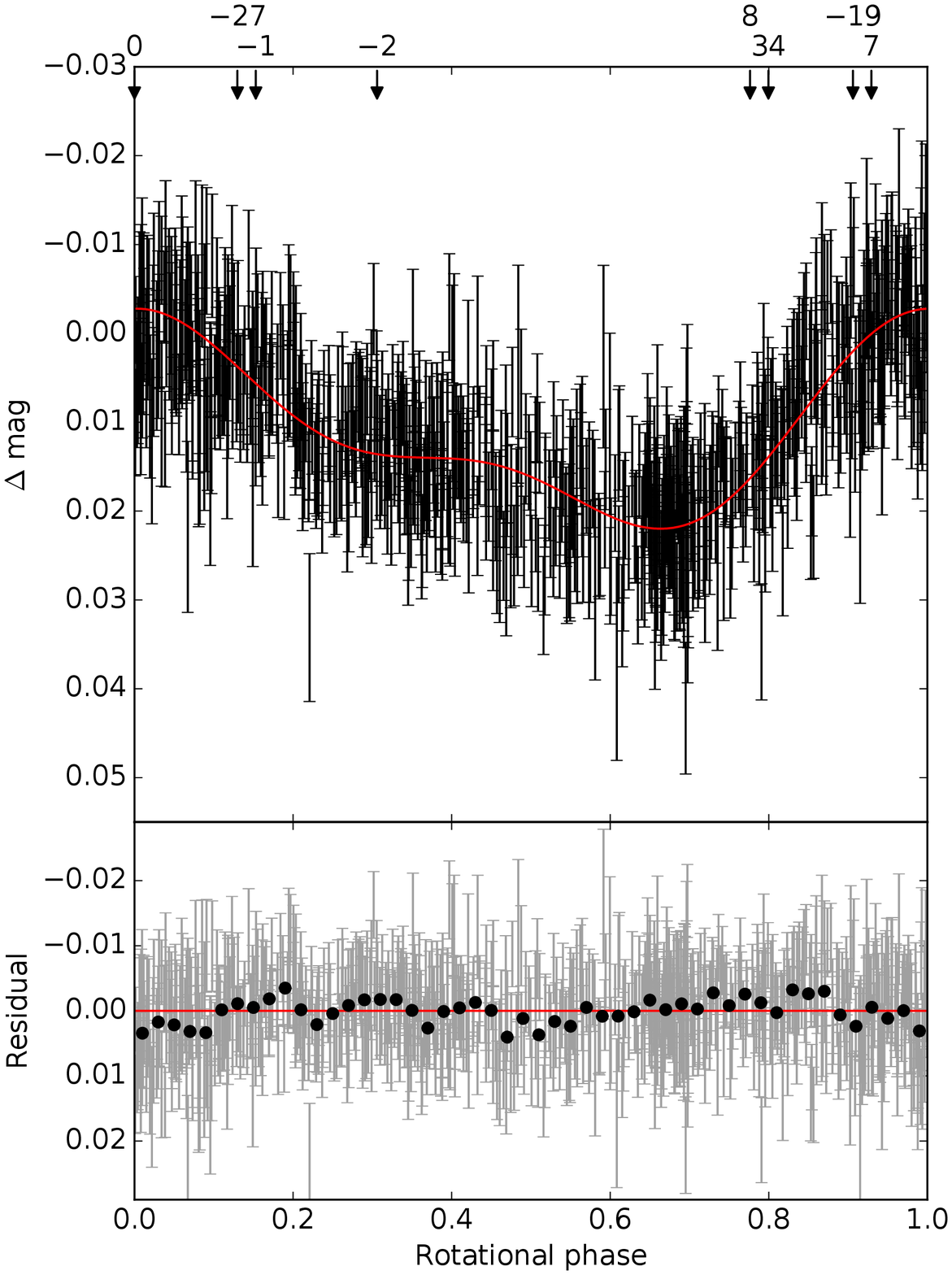}
\caption{Light curves for LP~261-75 with the best fit overlaid.  
  Left panels: primary eclipse windows with the out-of-eclipse
  modulation and magnitude zero point offsets removed to flatten the
  baseline.  Different eclipses are offset vertically for clarity, and
  the cycle number (integer part of the normalized orbital phase) is
  given on the right.  Right panels: out-of-eclipse modulation with
  the eclipse windows removed.  Only the instrumental effects
  (magnitude zero point offsets and common mode) were corrected.  The
  phases where eclipses were observed are indicated with arrows at the
  top of the diagram, giving the appropriate cycle numbers.  When
  plotting the residuals, the original data are shown in grey, and
  black points show the same data binned in phase into 50 equal size
  bins.}
\label{lp261_75_lcfit}
\end{figure*}

\begin{figure}
\centering
\includegraphics[width=3.5in]{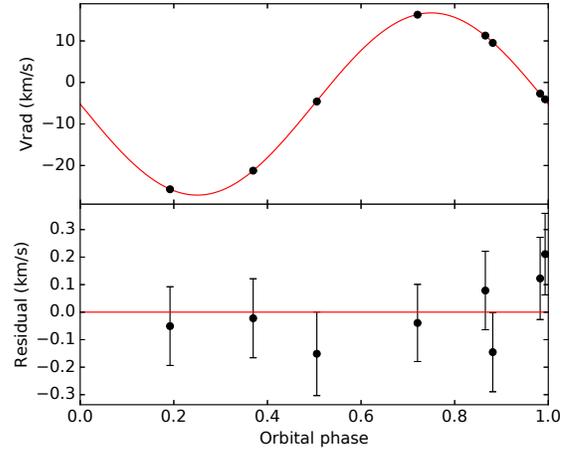}
\caption{Single-lined radial velocity orbit for LP~261-75 with the
  best fit overlaid (top) and residuals (bottom).}
\label{lp261_75_vrad}
\end{figure}

\begin{figure}
\centering
\includegraphics[width=3.5in]{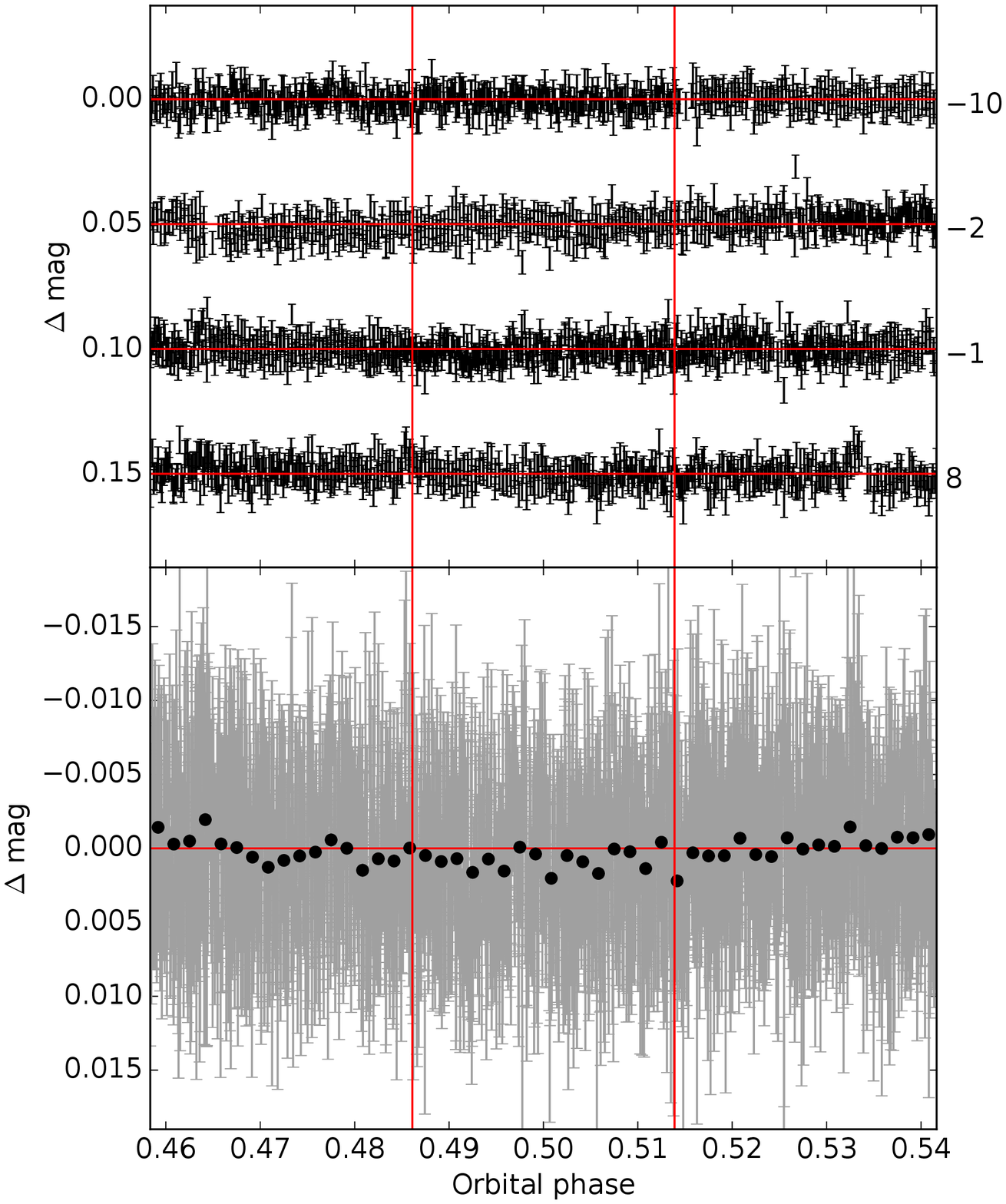}
\caption{Secondary eclipse window for LP~261-75.  Top panel:
  individual light curves with baseline flattened as Figure
  \ref{lp261_75_lcfit}, with the secondary eclipse duration indicated
  by vertical red lines.  Bottom panel: combined, phase-folded light
  curve, with bins as described for Figure \ref{lp261_75_lcfit}.  No
  secondary eclipse is detected.}
\label{lp261_75_sec}
\end{figure}

Table \ref{lp261_75_params} gives our orbital solution for LP~261-75,
and Figures \ref{lp261_75_lcfit} and \ref{lp261_75_vrad} show this
model overplotted on the data used in the analysis.  We also observed
several secondary eclipse windows which were not included in the
models, but are shown in Figure \ref{lp261_75_sec} to justify the
choice of fixing $J = 0$ in the solution.

\begin{deluxetable}{lr}
\tablecaption{\label{lp261_75_params} Parameters and uncertainties for LP~261-75.}
\tablecolumns{2}

\tablehead{
\colhead{Parameter}     & \colhead{Value}
}

\startdata
\hline
\multicolumn{2}{l}{Light curve jump parameters}\\
\hline
$R_2/R_1$                            & $0.29484 \pm 0.00034$ \\
$(R_1+R_2)/a$                        & $0.08843 \pm 0.00035$ \\
$\cos i$                             & $0.0152 \pm 0.0011$ \\
$F_1$                                & $0.84704 \pm 0.00019$ \\
$a_{11}$                             & $0.00299 \pm 0.00025$ \\
$b_{11}$                             & $0.00866 \pm 0.00025$ \\
$a_{12}$                             & $-0.00125 \pm 0.00019$ \\
$b_{12}$                             & $0.00396 \pm 0.00021$ \\
$C$                                  & $0.787 \pm 0.069$ \\
\hline
\multicolumn{2}{l}{Radial velocity jump parameters}\\
\hline
$\gamma$ (km/s)                      & $-5.193 \pm 0.054$ \\
$K_{1}$ (km/s)                       & $21.942 \pm 0.081$ \\
$s_{1}$ (km/s)                       & $0.125 \pm 0.037$ \\
\hline
\multicolumn{2}{l}{Derived parameters (MLR-independent)}\\
\hline
$i$ (deg)                            & $89.131 \pm 0.065$ \\
$e$ (95\% credible)                  & $< 0.007$ \\
$f_1(M)$ (${\rm M}_\odot$)           & $0.002060 \pm 0.000023$ \\
\hline
\multicolumn{2}{l}{Derived parameters (MLR-dependent)}\\
\hline
$q$                                  & $0.2166 \pm 0.0041$ \\
$M_1$ (${\rm M}_\odot$)              & $0.300 \pm 0.015$ \\
$M_2$ (${\rm M}_\odot$)              & $0.0650 \pm 0.0020$ \\
$R_1$ (${\rm R}_\odot$)              & $0.3131 \pm 0.0049$ \\
$R_2$ (${\rm R}_\odot$)              & $0.0923 \pm 0.0015$ \\
$v_{{\rm rot},1}$ (km/s)             & $7.13 \pm 0.11$ \\
\enddata

\end{deluxetable}

We caution that the primary eclipse light curves show evidence of
frequent spot crossings in the residuals, and the depths appear to be
somewhat variable, meaning our nominal value for the radius ratio may
exhibit systematic errors depending on the distribution of spots on
the stellar photosphere.  The stellar spin and orbital period are not
synchronized, which provides some information on the influence of the
asymmetric component of the spot distribution (e.g., as discussed in
\citealt{2011ApJ...742..123I}) but the number of eclipses available at
present is rather limited so we have not attempted such an analysis.

Additional trial solutions were run allowing a different radius ratio
for each of the $8$ primary eclipses to estimate the contribution of
this source of error.  The resulting unweighted mean of these radius
ratios was found to be compatible with the joint solution given in
Table \ref{lp261_75_params} but the empirical error in the mean was
$0.0012$, or $0.00064$ rejecting one outlier (the eclipse numbered $7$
in Fig. \ref{lp261_75_lcfit}).  We therefore suggest the uncertainty
reported for this parameter in the table may need to be inflated to
account for the effect of the spot crossings.

We further note that the two velocities close to orbital phase $1.0$
shown at the right-hand side of Figure \ref{lp261_75_vrad}
inadvertently overlapped the primary eclipse at the end of the
exposures, so could be influenced by the Rossiter-McLaughlin effect.
The resulting velocity anomaly would be approximately $+0.1\ \kms$ for
these data points if the system is spin-orbit aligned.  While they do
show slightly elevated positive residuals compared to the model, we do
not consider them to be significant at present.  The
Rossiter-McLaughlin effect has not been accounted for in modeling, and
while the estimated observational uncertainties (via the $s_{1}$
parameter) are inflated by the presence of these residuals it is still
possible some systematic errors exist in $\gamma$ and $K_{1}$ and any
parameters derived therefrom.

Since this is a single-lined spectroscopic binary, an estimate of the
primary mass is needed to derive the properties of the secondary.
LP~261-75 is a close kinematic match to the AB Dor moving group
(e.g., \citealt{2015ApJS..219...33G}), so we must
first assess the age of the system in order to determine which
relations are appropriate for estimating the primary mass.  Using our
value for the $\gamma$ velocity and the astrometric parameters in
Table \ref{photparams} we obtain $(U, V, W) = (-5.9 \pm 1.1, -28.7 \pm
1.3, -14.8 \pm 0.9)\ \kms$.  We quantify the kinematic match to AB Dor
using the BANYAN $\Sigma$ web tool \citep{2018ApJ...856...23G},
obtaining a membership probability of $99\%$.  However, there must
also be independent observational evidence of youth before an object
can be considered a moving group member, which we now proceed to examine.

This analysis is complicated by the structure of the AB Dor moving
group.  AB Dor ``stream'' stars do not appear to all have the same
age and chemical composition, with the population likely consisting of
a subsample of young stars which share a common origin with the main
AB Dor ``nucleus'', whereas the rest probably do not
(e.g., \citealt{2013ApJ...766....6B}).  The age of the nucleus and
young stream members of AB Dor is estimated to be 130--200 Myr in
the recent work of \citet{2015MNRAS.454..593B}.  Ages in this range
have also been suggested to explain the observed properties of the
M-dwarf primary in the present system, predominantly its high activity
level (e.g., \citealt{2006PASP..118..671R,2009ApJ...699..649S}), but it
is important to note that activity can be influenced by tides in close
spectroscopic binaries, and we further find that strong H$\alpha$
activity and rotation periods of a few days may persist beyond $1$ Gyr
in some mid-M systems
(e.g., \citealt{2011ApJ...727...56I,2016ApJ...821...93N}).

\begin{figure}
\centering
\includegraphics[width=3.5in]{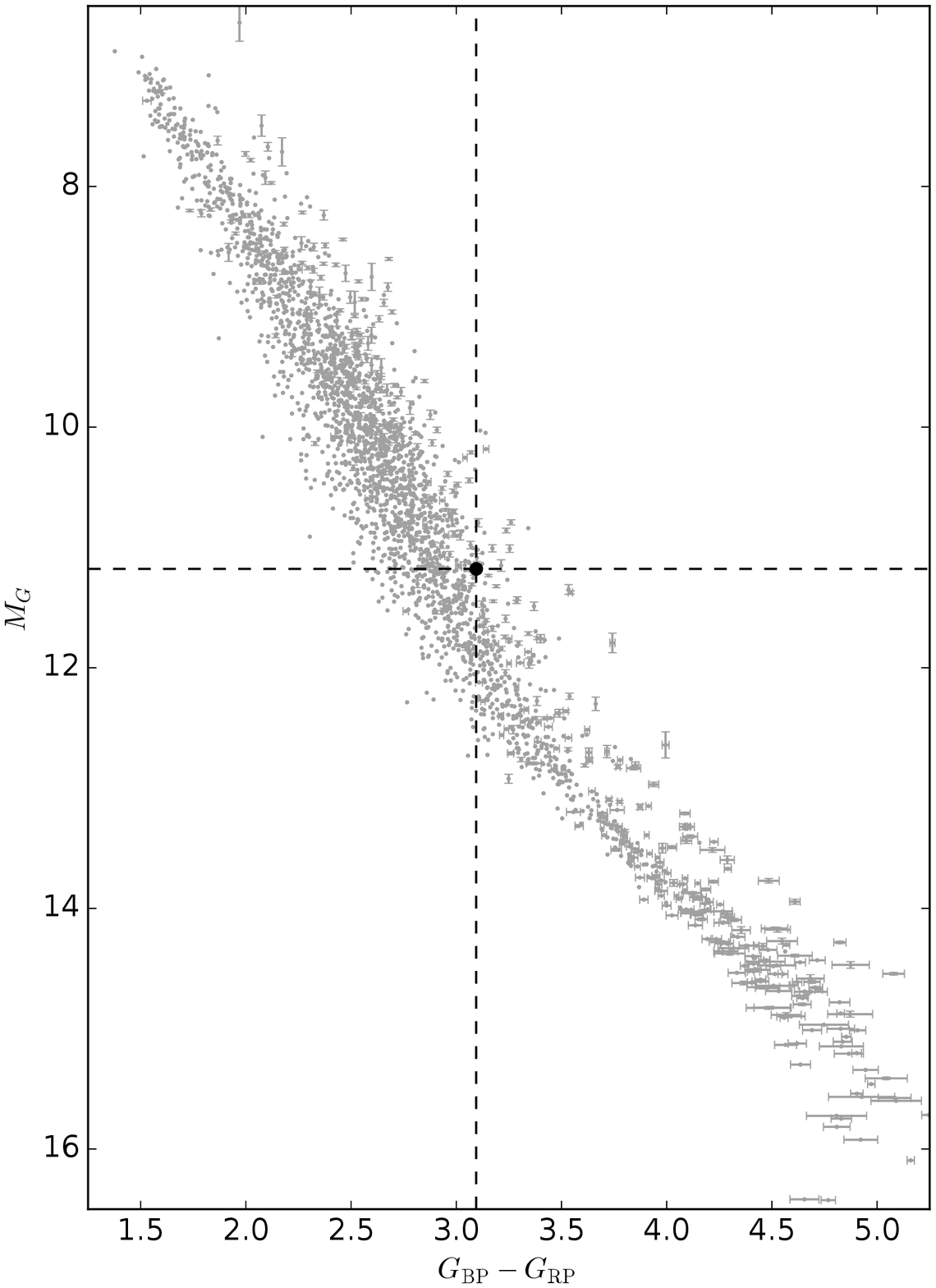}
\caption{$M_G$ versus $G_{\rm BP}-G_{\rm RP}$ color-magnitude diagram
  of the \citet{2008PASP..120..317N} parent sample for MEarth-North,
  using the 2MASS cross-match table provided in GAIA DR2 to retrieve
  the GAIA data, and excluding known unresolved multiples.  The
  criteria given in \citet{2018arXiv180409378G} were used to filter
  potentially contaminated $G_{\rm BP}$ and $G_{\rm RP}$ measurements
  based on the excess factor.  LP~261-75 is plotted in solid black,
  and the position is indicated by the dashed lines.}
\label{lp261_75_cmd}
\end{figure}

We do not find any clear observational evidence of youth in LP~261-75
at present.  The position of the system on an $M_G$ vs $G_{\rm
  BP}-G_{\rm RP}$ color-magnitude diagram (Figure \ref{lp261_75_cmd})
falls at the red end of the main locus of MEarth targets, which could
simply result from high metallicity (e.g.,
\citealt{2016ApJ...818..153D}).  The density of the primary is
constrained from the light curve analysis, but is not strongly
discriminating once we account for the known tendency of stellar
models to underpredict the radii of field stars.  The properties of
the secondary are more sensitive to age, with evolutionary models
\citep{2000ApJ...542..464C,2003A&A...402..701B} predicting the
secondary would be approximately $0.14-0.15\ \rsol$ and sufficiently
luminous to produce a secondary eclipse of several percent in MEarth
if it was in the $100-200$ Myr age range.  The observed properties of
the secondary and the lack of secondary eclipses are instead more
consistent with the model predictions for Gyr ages.

Based on this argument, we conclude that the system is likely
sufficiently old to apply relations for normal field stars to 
estimate the primary mass.  We therefore use the K-band
mass-luminosity relation (MLR) from \citet{2016AJ....152..141B} in
conjunction with the 2MASS K-band magnitude and parallax from Table
\ref{photparams}.  We assume an uncertainty of $0.09$ mag in absolute
magnitude on the MLR, as stated in Table 12 of
\citet{2016AJ....152..141B}, and use the double-exponential form
(absolute magnitude as a function of mass) which we find is better
behaved at the extremes of the mass range than the polynomial
relations.  The parameters depending on the adopted MLR are indicated
in Table \ref{lp261_75_params}.

\subsection{LP~796-24}

\begin{figure*}
\centering
\includegraphics[angle=270,width=6.0in]{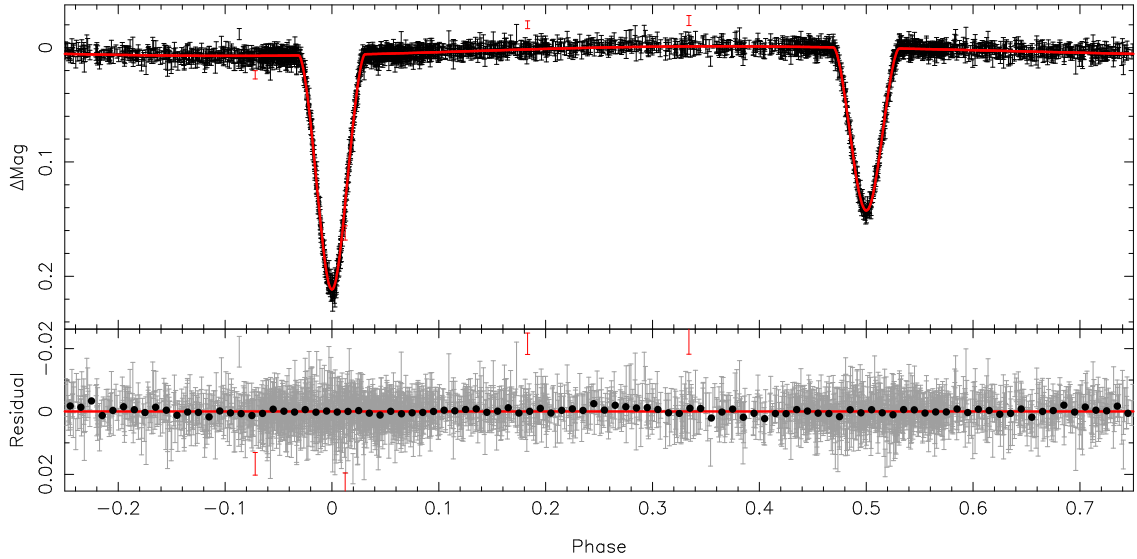}

\vspace{2ex}

\caption{Phase-folded light curve for LP~796-24 with best fit
  overlaid.  Panels are the same as described for Figure
  \ref{lp107_25_modelfit}.}
\label{lp796_24_lcfit}
\end{figure*}

Figure \ref{lp796_24_lcfit} shows the light curve for this system.
As discussed in \S \ref{vrad_sect}, the spectroscopic quantities are
not reliably measured, so we have not attempted a full solution and
provide only the orbital ephemeris determined from the MEarth light
curves in Table \ref{ephparams}.  Higher signal to noise ratio
spectra, reliably detecting all three components, would be needed for
a full analysis.

\subsection{LP~991-15}
\label{lp991_15_sect}

\begin{figure}
\centering
\includegraphics[angle=270,width=3.3in]{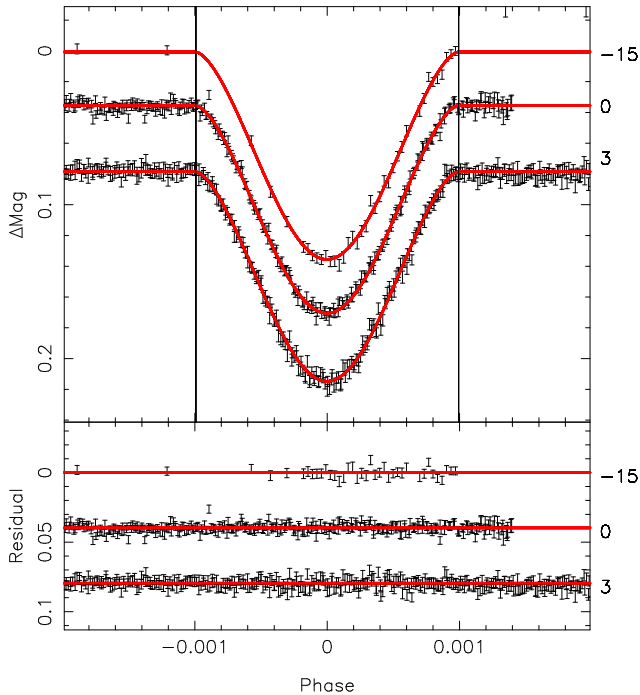}

\vspace{2ex}

\caption{Primary eclipses for LP~991-15.  Vertical offsets and light
  curve corrections are the same as described for Figure
  \ref{lp261_75_lcfit}.}
\label{lp991_15_pri}
\end{figure}

\begin{figure*}
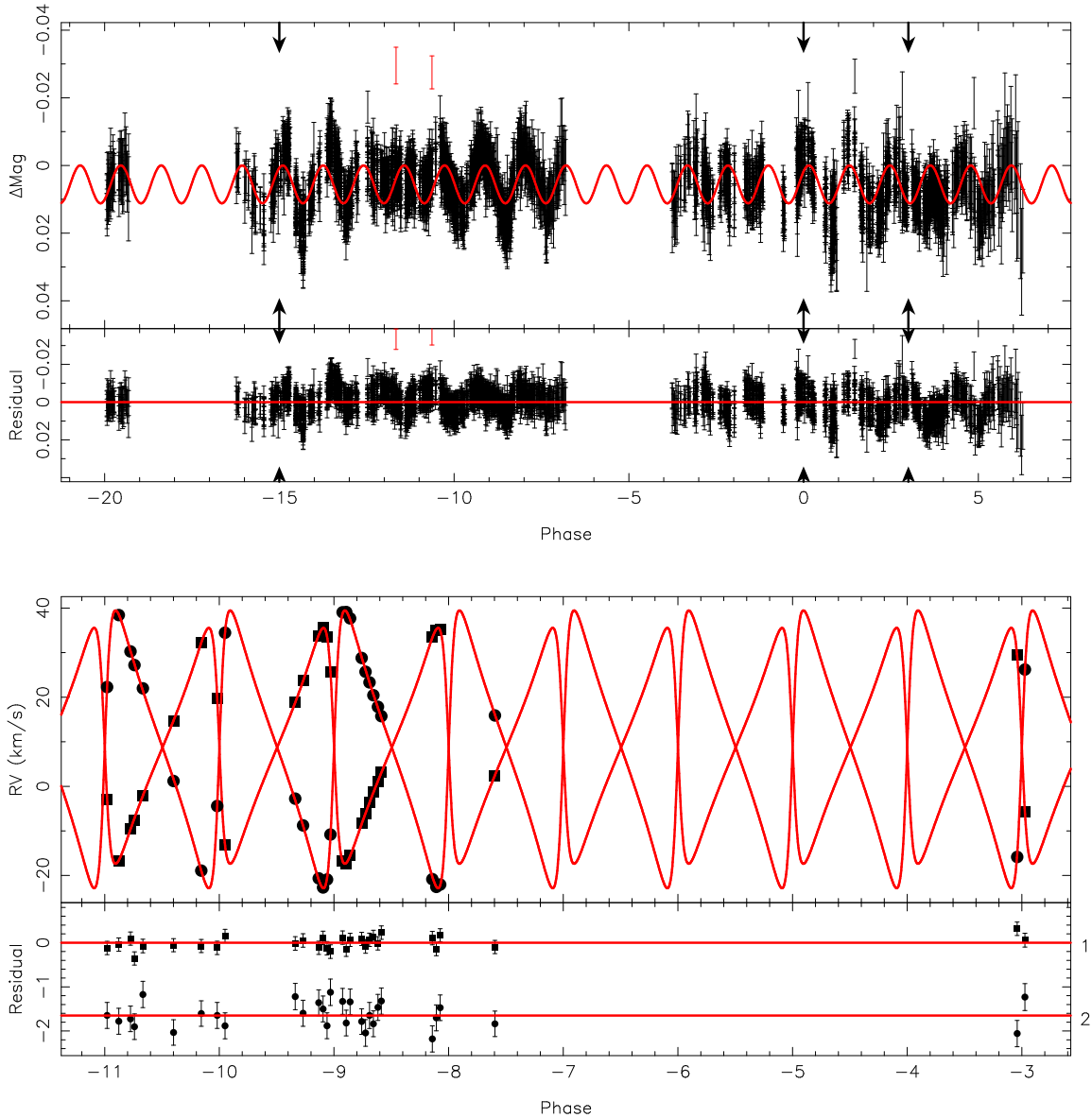

\centering
\includegraphics[angle=270,width=6in]{f9a.eps}

\vspace{4ex}

\includegraphics[angle=270,width=6in]{f9b.eps}

\vspace{2ex}

\caption{Top panels: out-of-eclipse light curve for LP~991-15, plotted
  as for Figure \ref{lp261_75_lcfit} with the observed primary eclipse
  epochs indicated by arrows.  Bottom panels: radial velocity curve.
  Symbols and vertical offsets are the same as for Figure 
  \ref{lp107_25_modelfit}.}
\label{lp991_15_ooe_vrad}
\end{figure*}

Figure \ref{lp991_15_pri} shows the eclipse light curves for this
system, and Figure \ref{lp991_15_ooe_vrad} shows the out-of-eclipse
modulation and the radial velocities.  The orbital solution is given
in Table \ref{lp991_15_params}.

\begin{deluxetable}{lr}
\tablecaption{\label{lp991_15_params} Parameters and uncertainties for LP~991-15.}
\tablecolumns{2}

\tablehead{
\colhead{Parameter}     & \colhead{Value}
}

\startdata
\hline
\multicolumn{2}{l}{Jump parameters}\\
\hline
$J$ (see note)                       & $0.93 \pm 0.12$ \\
$(R_1+R_2)/a$                        & $0.01455 \pm 0.00021$ \\
$\cos i$                             & $0.01957 \pm 0.00071$ \\
$e \cos \omega$                      & $0.0136 \pm 0.0011$ \\
$e \sin \omega$                      & $0.51645 \pm 0.00095$ \\
$q$                                  & $0.8489 \pm 0.0024$ \\
$F_1$                                & $0.86340 \pm 0.00056$ \\
$a_{11}$                             & $0.00622 \pm 0.00031$ \\
$b_{11}$                             & $0.00578 \pm 0.00031$ \\
$(K_1+K_2)$ (km/s)                   & $57.605 \pm 0.088$ \\
$\gamma$ (km/s)                      & $8.727 \pm 0.026$ \\
$C$                                  & $1.042 \pm 0.025$ \\
$s_{1}$ (km/s)                       & $0.143 \pm 0.020$ \\
$s_{2}$ (km/s)                       & $0.277 \pm 0.037$ \\
\hline
\multicolumn{2}{l}{Derived parameters}\\
\hline
$i$ (deg)                            & $88.878 \pm 0.041$ \\
$e$                                  & $0.51664 \pm 0.00096$ \\
$\omega$ (deg)                       & $88.49 \pm 0.13$ \\
$M_1$ ($\msol$)                      & $0.1969 \pm 0.0011$ \\
$M_2$ ($\msol$)                      & $0.16715 \pm 0.00072$ \\
\enddata

\tablecomments{The value of $J$ is determined by the combination of
  the priors on $R_2/R_1$ and $L_2/L_1$ and is not constrained
  observationally.  We state it only for completeness.}

\end{deluxetable}

Due to the lack of secondary eclipses, the light curve parameters in
this system depend on the assumptions (in particular, the
spectroscopic light ratio $\alpha$ and the adopted prior in the radius
ratio) to a greater degree than is usual for double-lined eclipsing
binaries with more normal configurations showing two eclipses.  The
surface brightness ratio parameter $J$, which is usually derived from
the relative depths of the primary and secondary eclipses, is largely
unconstrained by the data in this system, but it is still needed to
interpret the observed primary eclipse depth to extract $\cos i$ and
$(R_1+R_2)/a$.

While the radius ratio is not determined, in theory the sum of the
radii is still constrained by the observed eclipse duration.  In
practice, however, this inference also depends on other assumptions
such as limb darkening parameters and third light to an extent which
is not taken into account by the Monte Carlo procedure we have used to
estimate uncertainties, and we do not attempt to provide or
interpret this parameter given these difficulties.  This system is
therefore not currently useful to test the mass-radius relation.  It
is possible future highly precise light curves may be able to
alleviate some of these degeneracies in models of the primary
eclipse.

The component masses are well-determined, and while they depend on
$\cos i$ from the light curve solution, the mere existence of
primary eclipses is largely sufficient to constrain $\sin i$ for this
purpose given the long orbital period.  We caution that it is likely
the uncertainties in the masses stated in Table \ref{lp991_15_params}
have been underestimated due to neglecting correlated noise in the
radial velocity analysis, where the residuals in Figure
\ref{lp991_15_ooe_vrad} do appear to be correlated at a level
approximately equal to the uncertainties, but we suspect they are
indeed determined to better than $2\%$.  Combined with the astrometric
parallax, this system could therefore potentially be used to calibrate
the mass-luminosity or absolute magnitude relations.

\acknowledgments We thank Guillermo Torres and Eric Mamajek for
discussing details of the solutions and the kinematics, respectively,
and Todd Henry for encouragement to complete the CHIRON analysis.  We
also thank the staff at Fred Lawrence Whipple Observatory and Cerro Tololo
Inter-American Observatory for assistance in the construction and
operation of MEarth-North and MEarth-South; the CTIO 1.5m queue
observers: Carlos Corco, Alberto Miranda, Leonardo Paredes, and
Jacqueline Ser\'on, for assistance with gathering the CHIRON
observations; and the SMARTS queue managers: Emily MacPherson and
Imran Hasan for assistance in planning, scheduling, and reducing the
CHIRON observations.  The MEarth team acknowledges funding from the
David and Lucile Packard Fellowship for Science and Engineering
(awarded to D.C.).  This material is based on work supported by the
National Science Foundation under grants AST-0807690, AST-1109468,
AST-1004488 (Alan T. Waterman Award) and AST-1616624.  This
publication was made possible through the support of a grant from the
John Templeton Foundation.  The opinions expressed in this publication
are those of the authors and do not necessarily reflect the views of
the John Templeton Foundation.

This research has made extensive use of data products from the Two
Micron All Sky Survey, which is a joint project of the University of
Massachusetts and the Infrared Processing and Analysis Center /
California Institute of Technology, funded by NASA and the NSF, NASA's
Astrophysics Data System (ADS) bibliographic services, and the SIMBAD
and VizieR databases, operated at CDS, Strasbourg, France.  The
Digitized Sky Surveys were produced at the Space Telescope Science
Institute under U.S. Government grant NAG W-2166. The images of these
surveys are based on photographic data obtained using the Oschin
Schmidt Telescope on Palomar Mountain and the UK Schmidt
Telescope.  The plates were processed into the present compressed
digital form with the permission of these institutions.

{\it Facilities:} \facility{CTIO:1.5m (CHIRON)}, \facility{FLWO:1.5m (TRES)}

\bibliography{references}

\end{document}